\documentclass[sigconf]{acmart}

\usepackage{subcaption}
\usepackage{graphicx}
\usepackage{dblfloatfix}
\usepackage{algorithm}
\usepackage{algorithmic}   
\usepackage{tikz}

\newcommand{\name}{Camel}
\newcommand{\sampler}{Bandwidth Sampler}

\AtBeginDocument{%
  }

\copyrightyear{2026}
\acmYear{2026}
\setcopyright{cc}
\setcctype{by-nc-nd}
\acmConference[WWW '26]{Proceedings of the ACM Web Conference 2026}{April 13--17, 2026}{Dubai, United Arab Emirates}
\acmBooktitle{Proceedings of the ACM Web Conference 2026 (WWW '26), April 13--17, 2026, Dubai, United Arab Emirates}
\acmPrice{}
\acmDOI{10.1145/3774904.3792535}
\acmISBN{979-8-4007-2307-0/2026/04}

\begin{document}

\title{\name: Frame-Level Bandwidth Estimation for Low-Latency Live Streaming under Video Bitrate Undershooting}


\settopmatter{authorsperrow=5}
\author{Liming Liu}
\author{Zhidong Jia}
\author{Li Jiang}
\affiliation{%
  \institution{Peking University}
  \country{} 
}

\author{Wei Zhang}
\author{Lan Xie}
\author{Feng Qian}
\affiliation{%
  \institution{ByteDance Ltd.}
  \country{}
}

\author{Leju Yan}
\author{Bing Yan}
\author{Qiang Ma}
\affiliation{%
  \institution{ByteDance Ltd.}
  \country{}
}

\author{Zhou Sha}
\author{Wei Yang}
\author{Yixuan Ban}
\affiliation{%
  \institution{ByteDance Ltd.}
  \country{}
}

\author{Xinggong Zhang}
\affiliation{%
  \institution{Peking University}
  \country{}
}

\renewcommand{\shortauthors}{Liming Liu et al.}

\begin{abstract}
Low-latency live streaming (LLS) has emerged as a popular web application, with many platforms adopting real-time protocols such as WebRTC to minimize end-to-end latency. However, we observe a counter-intuitive phenomenon: even when the actual encoded bitrate does not fully utilize the available bandwidth, stalling events remain frequent. This insufficient bandwidth utilization arises from the intrinsic temporal variations of real-time video encoding, which cause conventional packet-level congestion control algorithms to misestimate available bandwidth. When a high-bitrate frame is suddenly produced, sending at the wrong rate can either trigger packet loss or increase queueing delay, resulting in playback stalls.

To address these issues, we present \name, a novel frame-level congestion control algorithm (CCA) tailored for LLS. Our insight is to use frame-level network feedback to capture the true network capacity, immune to the irregular sending pattern caused by encoding. \name\ comprises three key modules: the \textbf{Bandwidth and Delay Estimator} and the \textbf{Congestion Detector}, which jointly determine the average sending rate, and the \textbf{Bursting Length Controller}, which governs the emission pattern to prevent packet loss. 

We evaluate \name\ on both large-scale real-world deployments and controlled simulations. In the real-world platform with 250M users and 2B sessions across 150+ countries, \name\ achieves up to a 70.8\% increase in 1080P resolution ratio, a 14.4\% increase in media bitrate, and up to a 14.1\% reduction in stalling ratio. In simulations under undershooting, shallow buffers, and network jitter, \name\ outperforms existing congestion control algorithms, with up to 19.8\% higher bitrate, 93.0\% lower stalling ratio, and 23.9\% improvement in bandwidth estimation accuracy. 
\end{abstract}





\maketitle

\vspace{-4mm}
\section{Introduction}

As interactive live streaming has become a mainstream form of web-based content delivery~\cite{global-phenomena24}, the demand for low-latency live streaming (LLS) has surged, making it a pivotal technology in today’s web ecosystem~\cite{grandview2023asiapacific}. An end-to-end LLS system typically consists of two core components: the upstream transmission from broadcasters to the relay server, and the downstream distribution from the server to end-users. The streaming latency, playback smoothness, and video quality are the primary concerns for LLS. To reduce the latency, more and more LLS providers are switching the upstream link to real-time protocols such as WebRTC or SRT~\cite{halder2021fybrrstream,santos2016real,zhang2024harnessing}.

\begin{figure}[t]
  \centering
  \includegraphics[width=\linewidth]{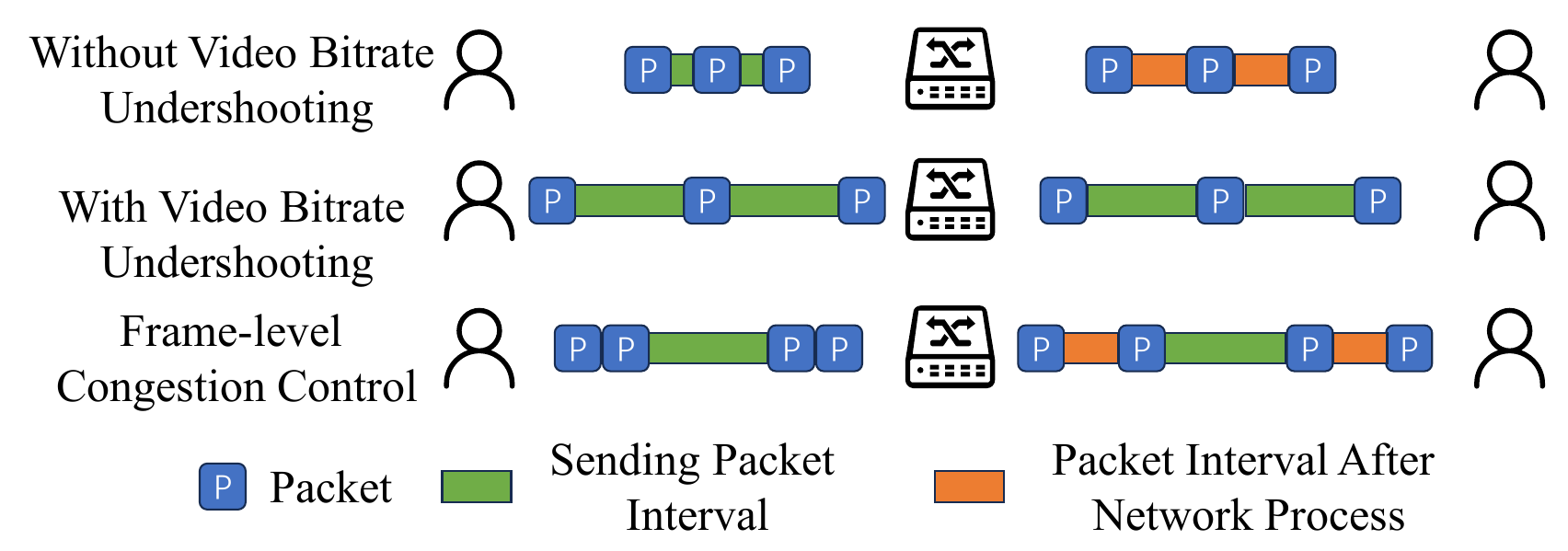}
  \vspace{-8mm}
  \caption{Without video bitrate undershooting, every packet can detect the network information; When video bitrate undershooting, feedback signals are also influenced by bitrate undershooting, causing estimation distortion; When using frame-level congestion control, some network information can be correctly measured within the individual frame burst.}
  \label{fig:intro:insight:cause:reason} 
  \vspace{-4mm}
\end{figure}


\begin{figure}[t]
  \centering

  \begin{tikzpicture}
     \node[anchor=south west,inner sep=0] (img) at (0,0) {
      \begin{minipage}[b]{0.48\linewidth}
        \includegraphics[width=\linewidth]{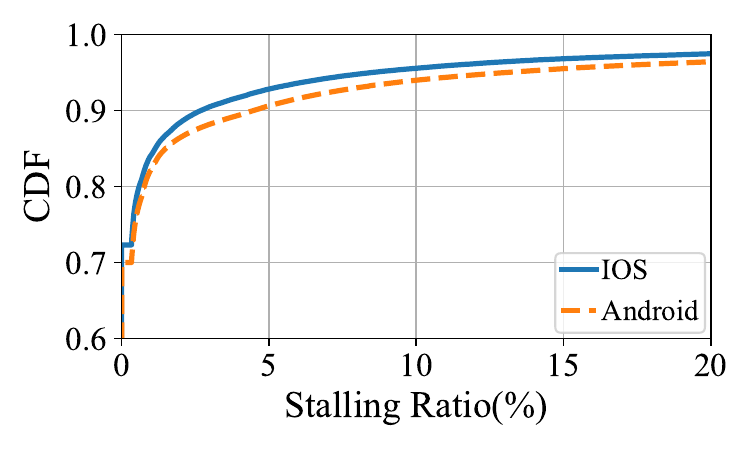}
      \end{minipage}
      \begin{minipage}[b]{0.48\linewidth}
        \includegraphics[width=\linewidth]{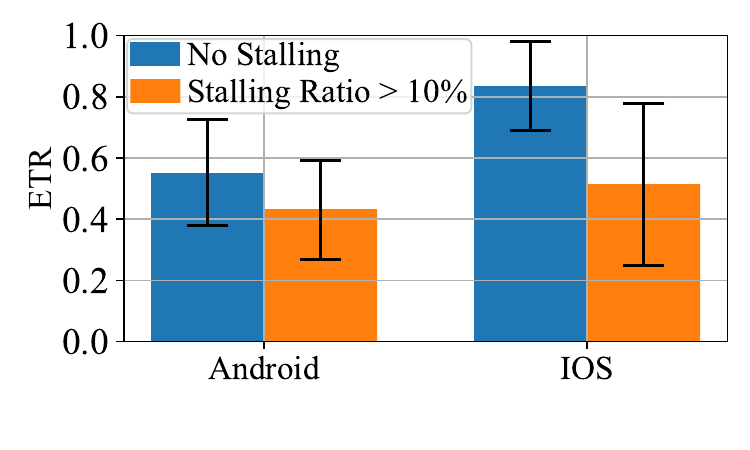}
      \end{minipage}
    };

    \begin{scope}[x={(img.south east)},y={(img.north west)}]
        \fill[white] (0.52,0.1) rectangle (0.546,0.9);

        \node[rotate=90, anchor=west, font={\scriptsize}] at (0.529,0.18) {Bitrate / Bandwidth};
    \end{scope}

  \end{tikzpicture}
  \vspace{-4mm}
  \caption{Distribution of stalling ratio and uplink bandwidth utilization among live streams' uploading links. Most live streams only utilize a moderate portion of the estimated uplink bandwidth. But the stalling ratio remains high. Additionally, in traces where stalling occurs, the video bitrate is typically further below the estimated uplink bandwidth.}
  \label{fig:intro:etr_stall}
  \vspace{-5mm}
\end{figure}

However, we observed a surprisingly counter-intuitive phenomenon in one of the largest global video streaming platforms. As shown by real-world data across approximately 2 billion sessions in Figure \ref{fig:intro:etr_stall}, most uploaded live streams utilize only less than 80\% of the estimated uplink bandwidth. Even so, the stalling ratio remains high: over 10\% for more than 5\% of broadcasters. And more surprisingly, in traces where the stalling ratio exceeds 10\%, the video bitrate is typically further below the estimated bandwidth.

Why does the stalling ratio remain high even when bandwidth is not fully used? For convenience, we refer to the phenomenon where the actual sending bitrate is persistently lower than the target bitrate or the estimated available bandwidth as \textbf{undershooting}, and we measure its severity by the \textbf{R}atio of the video \textbf{E}ncoded-bitrate to its \textbf{T}arget-bitrate (ETR), because the target bitrate is typically aligned with the estimated bandwidth. 

The reasoning chain is long but insightful. As shown in figure \ref{fig:intro:insight:cause:reason}, the root cause lies in the nature of RTC video traffic itself, which violates the fundamental assumption of continuous backlogging that most congestion control algorithms rely on. When the sender always has enough packets waiting to be transmitted, the measured network signals are influenced only by the network conditions, and thus accurately reflect the available bandwidth. However, real-time encoders operate on a frame-by-frame basis, with frame sizes varying significantly over time depending on content complexity and frame types. 
While encoders aim to match the average bitrate to the estimated bandwidth rather than sustaining a constant high rate, their instantaneous bitrate fluctuations often cause the sender to frequently drain the sending buffer.
This leads to bitrate undershooting that distorts network feedback signals: observed packet loss rates and inter-arrival intervals are no longer purely reflective of network capacity, but are also shaped by application-layer dynamics. Such distortion results in inaccurate bandwidth estimation. When the encoder suddenly generates a burst of high-bitrate frames, the wrong bandwidth estimation could have significant consequences. If the sender transmits too aggressively, packets are dropped and transmission delays accumulate. If it transmits too conservatively, queueing delay increases. In both cases, once the delay exceeds the playback buffer, it leads to visible stalling events.

So can we design a congestion control mechanism that not only aligns with the encoder’s natural and irregular frame-level burst pattern, but also leverages this pattern to more accurately estimate the available bandwidth? The answer is yes. \textbf{Our key idea is to transmit video data in a short, bursty pattern rather than a continuous stream, then leverage \emph{frame-level network feedback} to capture the true network capacity, which is immune to the irregular inter-frame sending fluctuations. }

While promising, frame-level congestion control also introduces several unique challenges. First, identifying reliable frame-level indicators and accurately mapping them to bandwidth estimates under undershooting conditions is non-trivial. Second, detecting congestion at the frame level is difficult because delays of adjacent frames can interfere with each other. Third, controlling the length of each transmission burst is critical. Longer bursts provide more samples for stable bandwidth estimation, but excessive bursting can overwhelm small network buffers, causing packet loss. 

To address these challenges, we present \name, a novel frame-level congestion control algorithm tailored for LLS upstream video transmission. \name~consists of three key modules: the \textbf{Bandwidth and Delay Estimator} and the \textbf{Congestion Detector} jointly determine the average sending rate, while the \textbf{Bursting Length Controller} governs the packet emission pattern to avoid packet loss.

It is worth mentioning that \name\ is motivated by the packet train concept and leverages a similar methodology to estimate network bandwidth and achieve frame-level congestion control. However, unlike prior studies on packet train~\cite{ray2022sqp,wang2024pudica,fouladi2018salsify,jia2024burstrtc,flach2016internet,spang2023sammy,vokkarane2002burst,wei2006tcp,carrysignals,jain1986packet,dovrolis2004packet,keshav1991control,partridge2002swifter}, \name\ introduces a novel congestion control algorithm tailored for LLS. In particular, it proposes the inflight-delay congestion detection method (\S \ref{sec:design:cd}), which effectively reduces congestion misjudgments, and designs a bursting length controller (\S \ref{sec:design:blc}) to prevent packet loss that introduced by frame-level bursting. It also proposes a bandwidth indicator in \S \ref{sec:design:bdp-est}. 

We have conducted extensive testing of our system in both large-scale real-world scenarios and controlled simulated environments. In real-world tests, our system was deployed on one of the largest live streaming platform, serving 250M users and 2B sessions across 150+ countries, demonstrating up to a 70.8\% increase in 1080P resolution ratio, a 14.4\% increase in media bitrate and up to a 14.1\% reduction in stalling ratio. In simulated environments, we evaluated the algorithm's performance under video bitrate undershooting, shallow buffer and network jitter conditions. The results showed \name\ significantly outperforms existing CCAs, achieving up to a 19.8\% increase in bitrate, a 93.0\% reduction in stalling ratio, and a bandwidth estimation accuracy improvement of up to 23.9\%. 

The remainder of this paper is structured as follows. In Section~\ref{sec:background}, we present the background and motivation behind our work. Section~\ref{sec:overview} provides an overview of \name, including its key components and data flow. The detailed design of \name\ is described in Section~\ref{sec:design}. We evaluate our system in Section~\ref{sec:evaluation}, followed by a discussion of related work in Section~\ref{sec:related}. Finally, Section~\ref{sec:conclusion} concludes the paper.

\vspace{-2mm}
\section{Background and Motivation}
\label{sec:background}

\begin{figure}[t]
  \centering  
  \begin{subfigure}[b]{0.48\linewidth}
    \centering  
    \includegraphics[width=\linewidth]{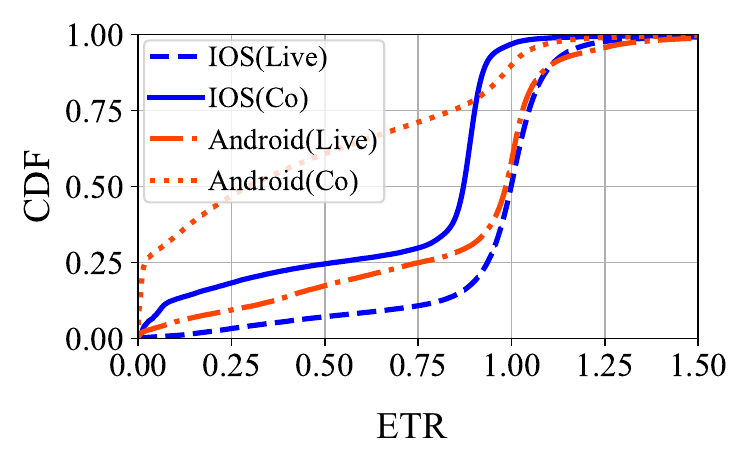}
    \vspace{-6mm}
    \subcaption{Distribution of actual-to-estimated bitrate ratio. } 
    \label{fig:back_app_limited_bw:a} 
  \end{subfigure}
  \hfill 
  \begin{subfigure}[b]{0.48\linewidth}
    \centering  
    \includegraphics[width=\linewidth]{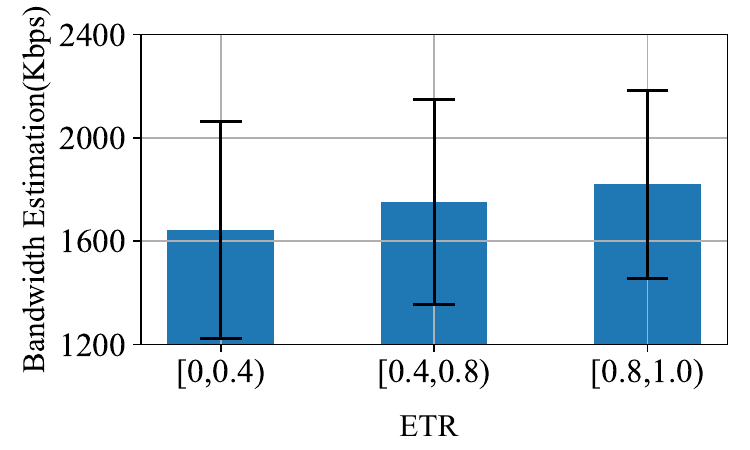} 
    \vspace{-6mm}
    \subcaption{Estimated bandwidth under different undershooting levels.} 
    \label{fig:back_app_limited_bw:b}
  \end{subfigure}
  \vspace{-2mm}
  \caption{Distribution and impact of undershooting in RTC-based upstream video streams.}
  \label{fig:back_app_limited_bw}
  \vspace{-6mm}
\end{figure}

\begin{figure}[t]
  \centering
  \begin{minipage}[b]{0.49\linewidth}
    \centering
    \includegraphics[width=\linewidth]{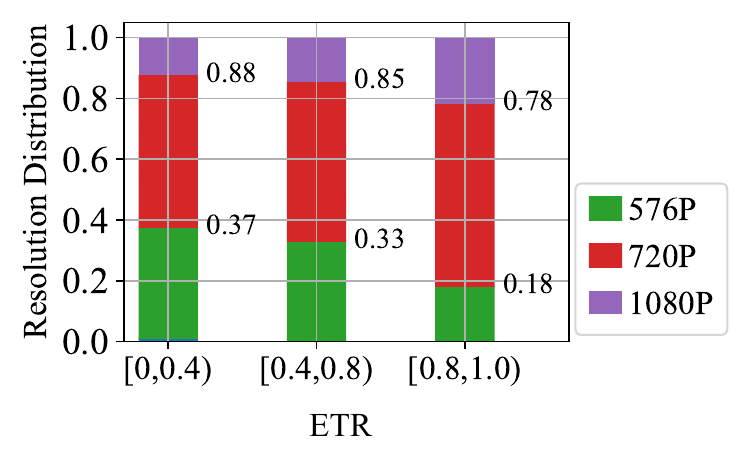}
    \vspace{-8mm}
    \caption{Lower ETR always leads to lower resolution.}
    \label{fig:back_app_res}
  \end{minipage}
  \hfill
  \begin{minipage}[b]{0.49\linewidth}
    \centering
    \includegraphics[width=\linewidth]{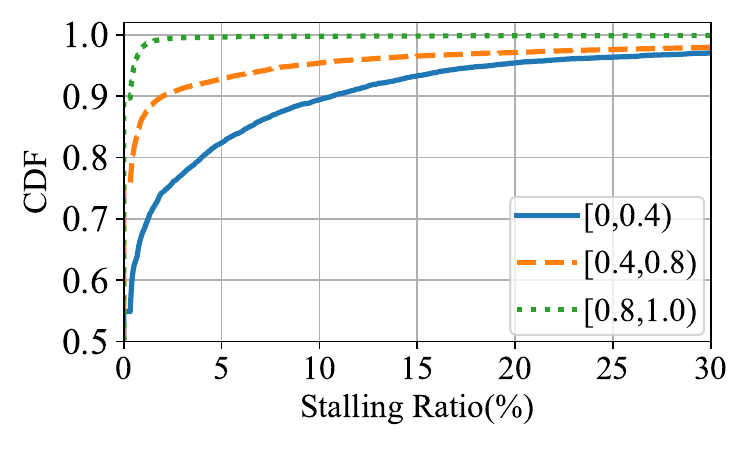}
    \vspace{-8mm}
    \caption{Lower ETR always leads to higher stalling ratio.}
    \label{fig:back_app_stall}
  \end{minipage}
  \vspace{-6mm}
\end{figure}

\subsection{Common Undershooting Causes Low QoE}

As shown by real-world data collected from one of the largest global video streaming platforms, we observe that the \emph{undershooting} behavior, where the actual sending bitrate is persistently lower than the target bitrate or the estimated available bandwidth, is widespread across RTC-based upstream sessions. Figure \ref{fig:back_app_limited_bw}(a) presents the distribution of the ratio between the actual encoded bitrate and the user’s estimated upload bandwidth within 12-second intervals. The results show that more than 50\% of streams exhibit varying degrees of undershooting. This common phenomenon can be attributed to several factors inherent to real-time video encoding. First, real-time encoders typically aim to match the average bitrate to the estimated available bandwidth rather than sustaining a constant high rate, since video content and frame type exhibits significant temporal fluctuations~\cite{huang2025ace}: I-frames can be more than 10 times larger than P or B-frames, leaving the encoder with unused bandwidth for most non-I frames. Second, to maintain low latency and prevent excessive queuing, real-time encoders frequently lower the instantaneous bitrate, further contributing to persistent undershooting~\cite{halder2021fybrrstream}.

To understand why undershooting matters, we next examine its impact on bandwidth estimation. Figure \ref{fig:back_app_limited_bw}(b) shows the estimated bandwidth across LLS streams grouped by different bitrate-to-bandwidth ratios. Since video bitrate ranges differ across resolutions, we use 720P streams as an example. The results indicate that as undershooting becomes more severe, existing congestion control algorithms (CCAs) tend to estimate bandwidth with a lower mean and higher variance. Given that the encoding bitrate is an inherent property of the video encoder, independent of transport-layer estimation~\cite{zhou2019learning}, this correlation suggests that undershooting directly suppresses the sender’s ability to probe for available capacity, leading to systematic bandwidth underestimation.

Such bandwidth underestimation has two direct consequences that jointly degrade user Quality of Experience (QoE). Figure \ref{fig:back_app_res} shows that the proportion of 720P+ videos drops significantly under lower bitrate-to-bandwidth ratios, while Figure \ref{fig:back_app_stall} shows that the stalling ratio consistently increases in these cases. Together, these findings demonstrate that undershooting not only pervasively exists but also critically undermines both playback smoothness and perceived video quality in LLS systems.
The reason is that, it prevents the sender from correctly pacing large video frames. When the encoder suddenly generates a high-bitrate frame, the sender faces two suboptimal choices. If it sends too aggressively, packets are dropped by the network and transmission delays accumulate. If it sends too conservatively, queueing delay increases. In both cases, once the delay exceeds the playback buffer, it leads to visible stalling events. Besides, under a persistently underestimated bandwidth, the adaptive bitrate (ABR) algorithm is misled to select lower video resolutions, resulting in visibly degraded visual quality~\cite{thang2014evaluation}.

\vspace{-4mm}
\subsection{Insight: Frame-Level Burst}

The root cause of this unreliability lies in the temporal structure of RTC video traffic itself, which violates the fundamental assumption of continuous backlogging that most congestion control algorithms rely on. When the sender always has enough packets waiting to be transmitted, the measured network signals like throughput, delay, and loss, are influenced only by the network conditions. Traditional packet-level CCAs therefore perform effectively in bulk data transfer scenarios, such as file downloads. In these cases, the dynamics of the congestion window (\emph{cwnd}) and the feedback-based adjustments of algorithms like BBR~\cite{cardwell2017bbr}, CUBIC~\cite{ha2008cubic}, or Copa~\cite{arun2018copa} closely track the true available bandwidth. However, RTC traffic fundamentally breaks this assumption. Unlike file transfer or bulk streaming applications that always have enough packets to send, real-time encoders operate on a frame-by-frame basis. Each frame must be fully encoded before transmission, and frame sizes vary significantly over time depending on different content complexities and frame types. As a result, the sender transmits short, irregular bursts of packets followed by idle gaps. These encoder-induced silences frequently drain the sending buffer, causing bitrate undershooting that distorts network feedback signals: such observed packet loss rate and inter-arrival intervals are no longer purely reflective of network capacity, but also shaped by application-layer dynamics. 

\begin{figure}[t]
  \centering
  \includegraphics[width=\linewidth]{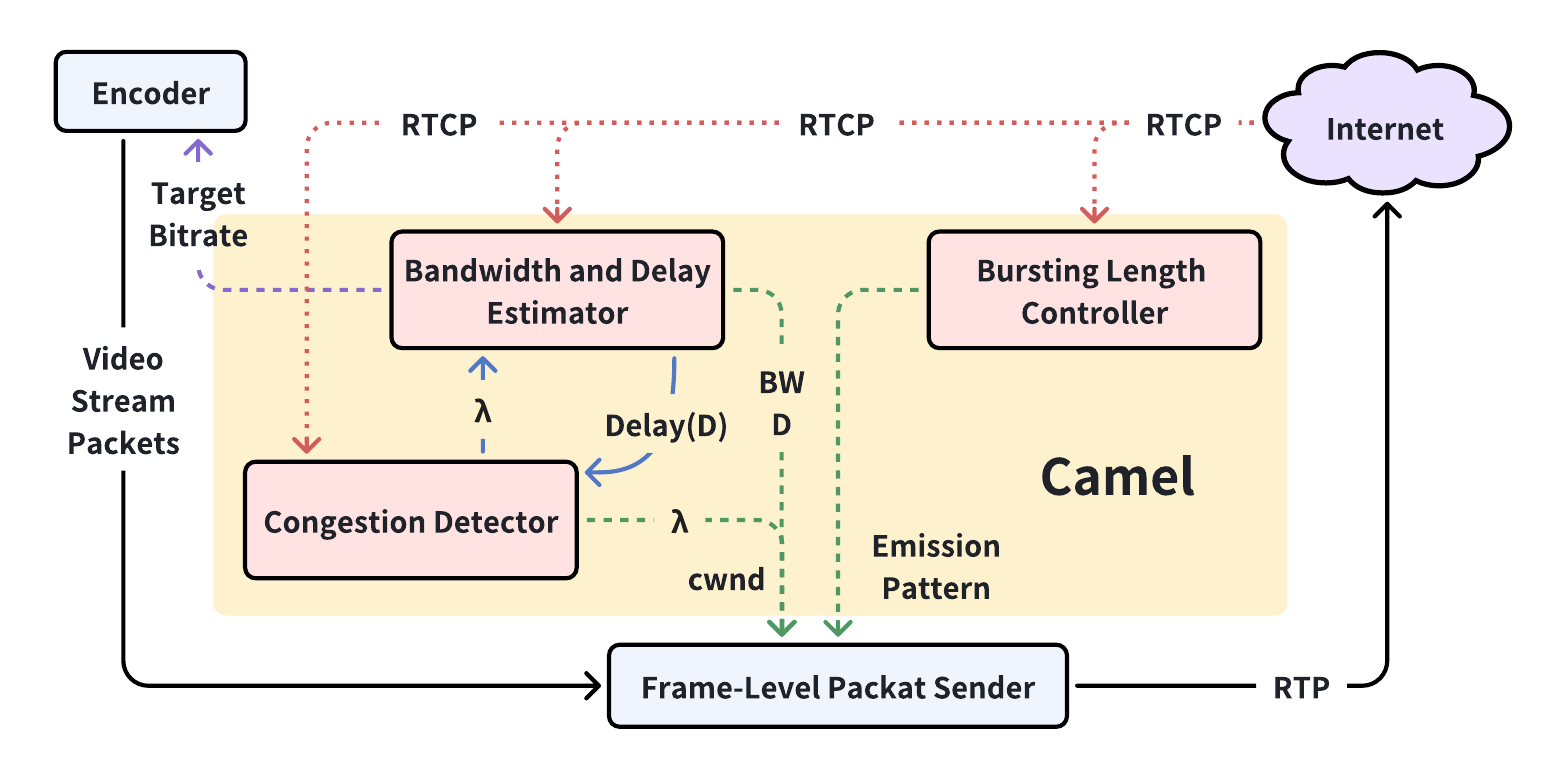}
  \vspace{-10mm}
  \caption{\name's system architecture.}
  \vspace{-4mm}
  \label{fig:design_arch} 
\end{figure}

Motivated by this observation, we ask: can we design a congestion control mechanism that not only aligns with the encoder’s natural and irregular frame-level burst pattern, but also leverages this pattern to more accurately estimate the available bandwidth?

Our answer is yes. As illustrated in Figure \ref{fig:intro:insight:cause:reason}, the key idea is to transmit video data in short, bursty segments rather than a continuous stream. We then leverage \emph{frame-level network feedback} which aggregates network signals within a frame to estimate the available bandwidth. By operating on frame-level statistics, the estimation becomes immune to the irregular sending pattern or idle gaps between bursts, enabling the congestion controller to accurately capture the true network capacity, even under undershooting. 

\begin{figure*}[ht]
  \centering
  \begin{subfigure}[b]{0.245\linewidth}
    \includegraphics[width=\linewidth]{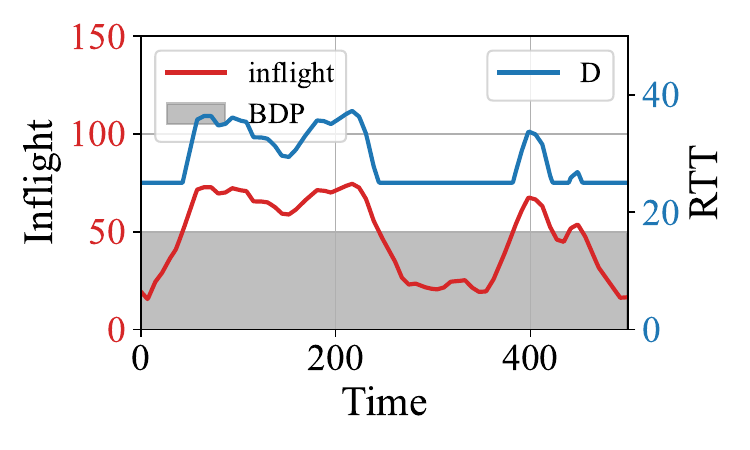}
    \vspace{-8mm}
    \caption{Flight and RTT trace.}
    \label{fig:design_cd_env}
  \end{subfigure}
  \hfill
  \begin{subfigure}[b]{0.245\linewidth}
    \includegraphics[width=\linewidth]{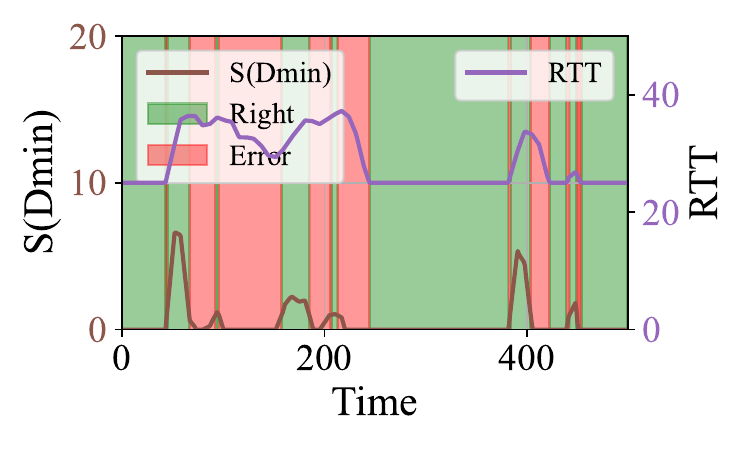}
    \vspace{-8mm}
    \caption{$S=D-Dmin$}
    \label{fig:design_cd_minRTT}
  \end{subfigure}
  \hfill
  \begin{subfigure}[b]{0.245\linewidth}
    \includegraphics[width=\linewidth]{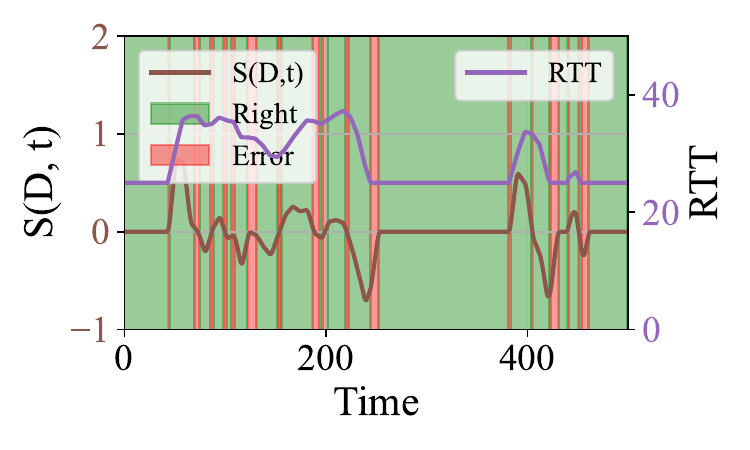}
    \vspace{-8mm}
    \caption{$S=\partial D / \partial t$}
    \label{fig:desgin_cd_RTT_t}
  \end{subfigure}
  \hfill
  \begin{subfigure}[b]{0.245\linewidth}
    \includegraphics[width=\linewidth]{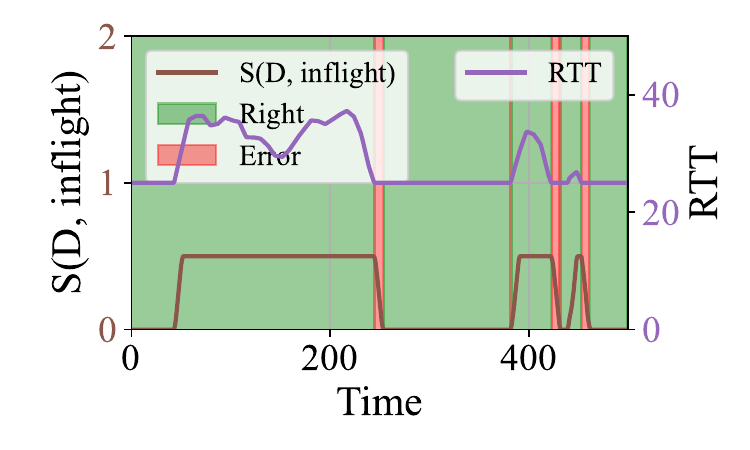}
    \vspace{-8mm}
    \caption{$S=\partial D / \partial \text{inflight}$}
    \label{fig:design_cd_RTT_inflight}
  \end{subfigure}
  \vspace{-4mm}
  \caption{The accuracy of different congestion signals, with $\partial D / \partial \text{inflight}$ achieves the most accurate result.}
  \label{fig:design_cd}
  \vspace{-4mm}
\end{figure*}

\subsection{Challenges}
\label{sec:challenges}

While frame-level CCA appears promising, it also introduces several unique challenges that must be carefully addressed.

\vspace{0.5em}
\noindent \textbf{Challenge 1: Bandwidth and Delay Estimation from Frame-Level Feedback.} 
Identifying reliable frame-level indicators and mapping them to accurate bandwidth estimates under undershooting conditions is a non-trivial task. The network feedback signals available at the frame level can be distorted by variations in frame size and encoding rate. Existing frame-level CCAs, such as SQP~\cite{ray2022sqp}, which relies on frame transport bandwidth, and Pudica~\cite{wang2024pudica}, which uses the bandwidth utilization ratio (BUR), are both affected by this issue. When a frame is smaller than the expected full-size frame, these algorithms compensate according to the undershooting ratio, which amplifies the statistical errors already present in small frames, leading to unstable or biased bandwidth estimation. 

\vspace{0.5em}
\noindent \textbf{Challenge 2: Congestion Detection Based on Frame-Level Delay.}  
Existing packet-level congestion signals have several drawbacks when applied to frame-level control. In particular, the delays of adjacent frames can easily interfere with each other, making it difficult to isolate congestion effects at the frame granularity. 

Loss-based algorithms~\cite{ha2008cubic} often suffer from low throughput in the presence of physical packet loss and fail to effectively control latency in networks with deep buffers.  
MinRTT-probing-based algorithms~\cite{cardwell2017bbr}, which use the minimum observed RTT as an estimate of propagation delay, are also unreliable in frame-level. A large tracking window for the minimum RTT cannot adapt to network jitter or dynamic background flows, while a small window risks producing non-convergent queueing delays once congestion arises.  
The delay gradient method~\cite{johnston2012webrtc}, adopted in GCC, is more resilient to network jitter. However, it cannot actively drain persistent queueing delay once it stabilizes. Moreover, time-ordered delay gradients depend on assumptions such as consistent packet sizes and fixed pacing rates, which do not hold in frame-level congestion control, where both the packet rate and frame size vary substantially.

\vspace{0.5em}
\noindent \textbf{Challenge 3: Bursting Length Control.}  
The buffer capacity of networks varies widely due to the heterogeneity of network infrastructures and carrier strategies~\cite{flach2016internet}. As a result, the length of each transmission burst must be carefully controlled. A longer bursting length allows the algorithm to incorporate more packet samples within each burst, improving the stability of bandwidth estimation. However, excessive bursting can easily overwhelm small buffers, causing packet losses that typically emerging near the tail. Therefore, it is crucial to adaptively configure the maximum bursting length according to the network environment, balancing the accuracy of bandwidth estimation against the risk of buffer overflow.
\vspace{-3mm}
\section{Overview of \name}
\label{sec:overview}

To address the above challenges, we present \name, a novel frame-level congestion control algorithm tailored for LLS upstream video transmission. A congestion controller must make two fundamental decisions: (1) how large the sending window (cwnd) should be, and (2) what packet emission pattern to use. Moreover, modern congestion control algorithms must ensure that the sending window closely matches the number of packets the network can accommodate, while rapidly responding to congestion signals~\cite{cardwell2017bbr,ha2008cubic}. So the sending window is typically set according to this estimate:

\vspace{-4mm}
\begin{equation}
    \mathrm{cwnd} = \gamma \cdot \hat{\mathrm{BDP}},
    \vspace{-1mm}
\end{equation}

\noindent where $\hat{\mathrm{BDP}}$ denotes the estimated bandwidth–delay product (BDP), and $\gamma$ is a dynamic scaling factor that adapts to observed congestion to trade off utilization and queuing delay.
In \name, we decompose these decisions into three modules: the \textbf{Bandwidth and Delay Estimator} (for $\hat{\mathrm{BDP}}$, \S \ref{sec:design:bdp-est}) and the \textbf{Congestion Detector} (for $\gamma$, \S \ref{sec:design:cd}) jointly determine the cwnd, while the \textbf{Bursting Length Controller} (\S \ref{sec:design:blc}) governs the packet emission pattern. Each module directly targets one of the challenges identified in \S ~\ref{sec:challenges}. 

Figure \ref{fig:design_arch} shows the overall architecture of \name: the \textbf{Bandwidth and Delay Estimator} continuously analyzes the incoming RTCP feedback to estimate both the available bandwidth and the delay. The estimated bandwidth is then mapped to the encoder’s target bitrate, guiding rate control at the application layer, and simultaneously passed to the \textbf{Packet Sender} module. The estimated delay is forwarded to the \textbf{Congestion Detector}, which combines delay trends and RTCP feedback to derive a congestion signal that reflects real-time network status. Meanwhile, the \textbf{Bursting Length Controller} determines the appropriate burst length for the next transmission period based on the estimated network conditions. Finally, the Packet Sender uses the outputs from all three modules (estimated bandwidth, delay, congestion state, and burst length) to schedule packet transmissions from the encoder in a pattern that aligns with frame-level bursts while respecting network constraints.

\begin{figure*}[ht]
  \centering
  \begin{subfigure}[b]{0.245\linewidth}
    \includegraphics[width=\linewidth]{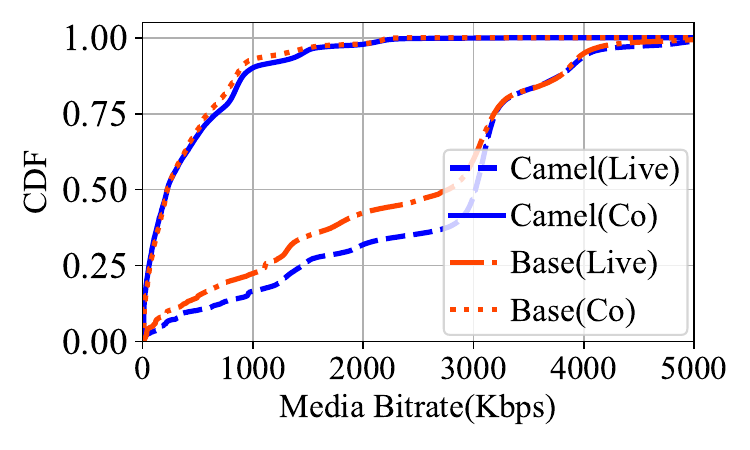}
    \vspace{-7mm}
    \caption{Media Bitrate (Android)}
    \label{fig:eval_ab_bitrate_android}
    \vspace{-1mm}
  \end{subfigure}
  \hfill
  \begin{subfigure}[b]{0.245\linewidth}
    \includegraphics[width=\linewidth]{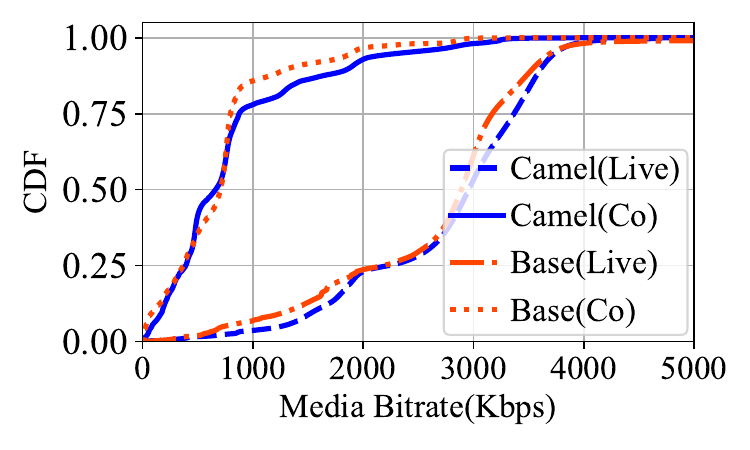}
    \vspace{-7mm}
    \caption{Media Bitrate (IOS)}
    \label{fig:eval_ab_bitrate_ios}
    \vspace{-1mm}
  \end{subfigure}
  \hfill
  \begin{subfigure}[b]{0.245\linewidth}
     \includegraphics[width=\linewidth]{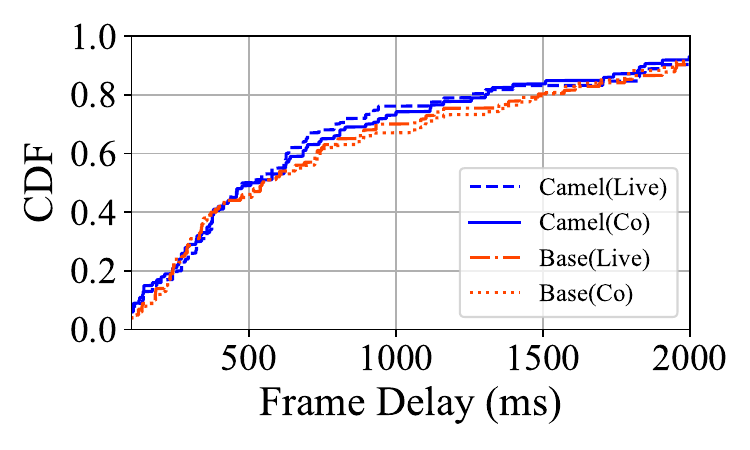}
     \vspace{-7mm}
    \caption{Frame Delay (Android)}
    \label{fig:eval_ab_delay_android}
    \vspace{-1mm}
  \end{subfigure}
  \hfill
  \begin{subfigure}[b]{0.245\linewidth}
    \includegraphics[width=\linewidth]{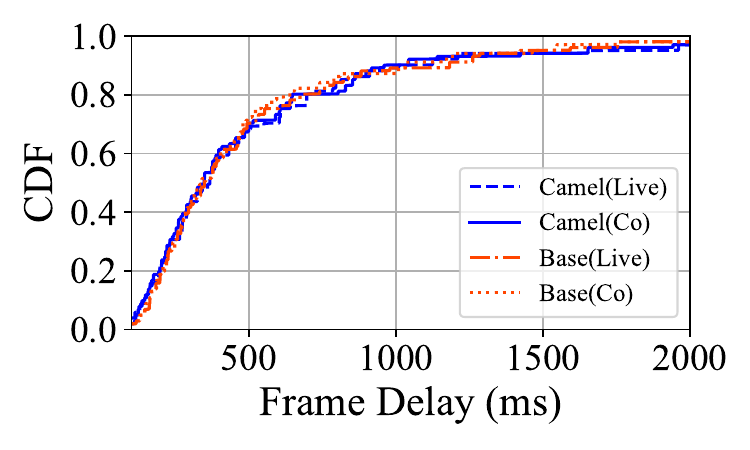}
    \vspace{-7mm}
    \caption{Frame Delay (IOS)}
    \label{fig:eval_ab_delay_ios}
    \vspace{-1mm}
  \end{subfigure}

    \begin{subfigure}[b]{0.245\linewidth}
     \includegraphics[width=\linewidth]{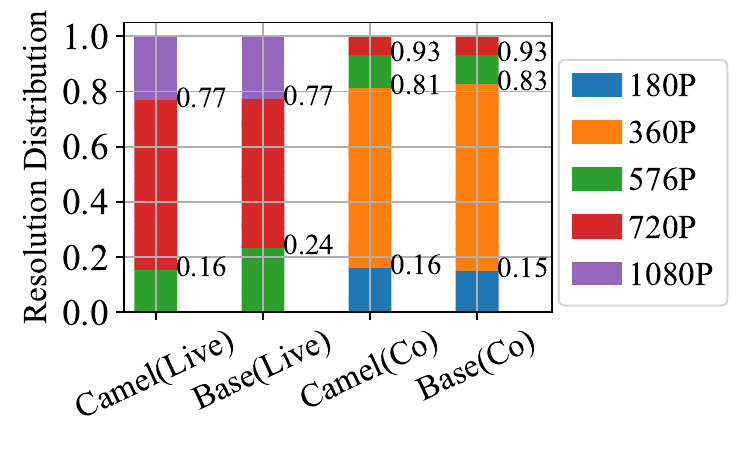}
     \vspace{-7mm}
    \caption{Resolution (Android)}
    \label{fig:eval_ab_res_android}
  \end{subfigure}
  \hfill
  \begin{subfigure}[b]{0.245\linewidth}
    \includegraphics[width=\linewidth]{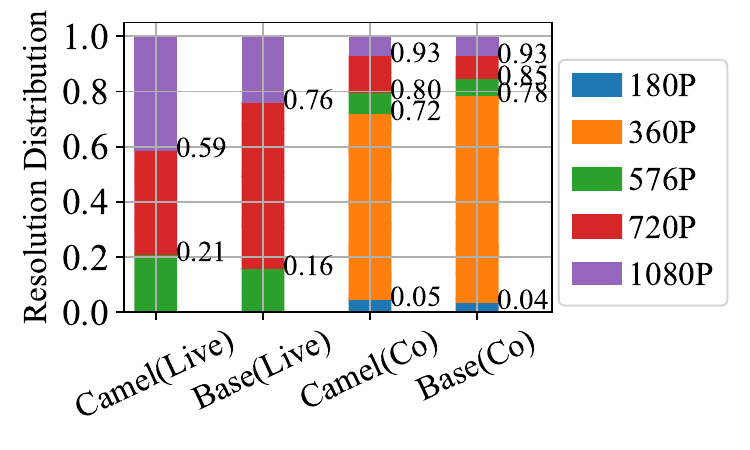}
    \vspace{-7mm}
    \caption{Resolution (IOS)}
    \label{fig:eval_ab_res_ios}
  \end{subfigure}
    \hfill
    \begin{subfigure}[b]{0.245\linewidth}
     \includegraphics[width=\linewidth]{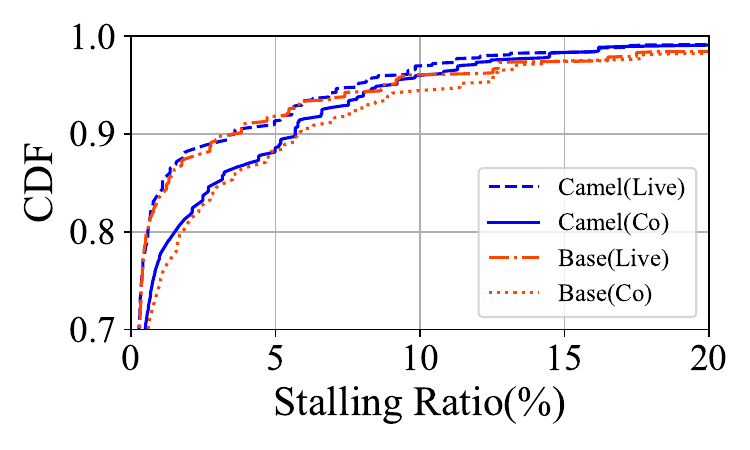}
     \vspace{-7mm}
    \caption{Stall Ratio (Android)}
    \label{fig:eval_ab_stall_android}
    \end{subfigure}
    \hfill
    \begin{subfigure}[b]{0.245\linewidth}
    \includegraphics[width=\linewidth]{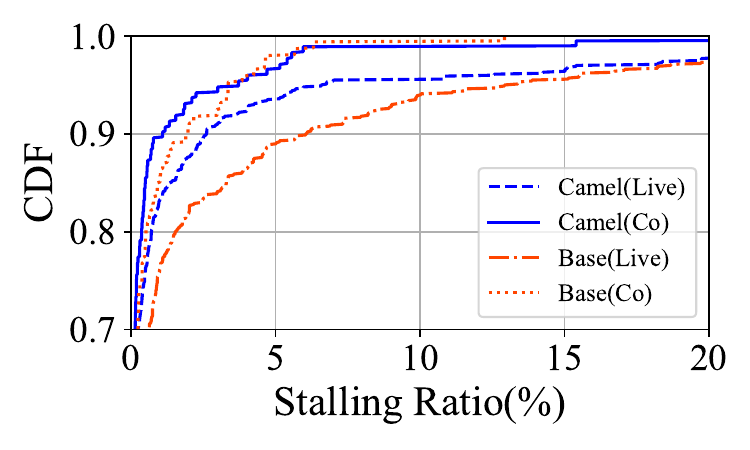}
    \vspace{-7mm}
    \caption{Stall Ratio (IOS)}
    \label{fig:eval_ab_stall_ios}
    \end{subfigure}

  \vspace{-4mm}
  \caption{Result of the large-scale A/B test, showing that \name~improves video QoE without compromising latency.}
  \label{fig:eva_ab_test_all}
  \vspace{-4mm}
\end{figure*}
\section{Design of \name}
\label{sec:design}

\subsection{Bandwidth and Delay Estimator}
\label{sec:design:bdp-est}
To address challenge 1, the bandwidth and delay estimator module takes RTCP feedback as input and outputs three key frame-level metrics: the bandwidth, the delay, and the bandwidth-delay product. 

\vspace{0.5em}
\noindent \textbf{Frame-Level Bandwidth Estimation.}
Inspired by the packet train algorithm~\cite{jain1986packet,flach2016internet} and Salsify~\cite{fouladi2018salsify}, our fundamental idea is that by transmitting the packets of each frame in rapid succession, minimizing the inter-packet intervals, the receiver can accurately infer the available network bandwidth from the arrival rate of these packets. For a frame consisting of $n$ packets, and these packets are numbered from $1$ to $n$, the frame-level bandwidth $B$ is defined as:

\vspace{-3mm}
\begin{equation}
    B = \frac{\sum_{i=2}^{n} S_i}{t_{recv_{n}} - t_{recv_{1}}},
    \label{equ:packet_train}
\end{equation}

\noindent where $S_i$ represents the size of the $i$-th packet and $t_{recv_{i}}$ denotes its receiving time. Since this bandwidth is calculated independently for each frame, the sender is relieved from maintaining continuous packet backlogging at the application layer.

\vspace{0.5em}
\noindent \textbf{Frame-Level Delay Estimation.}
To mitigate the self-introduced delay distortion caused by burst transmissions, we define the frame-level delay $D$ as the round-trip time (RTT) of the first packet in each frame. Unlike traditional pacing mechanisms that transmit packets at regular intervals, our frame-level approach sends all packets within a frame in a rapid burst. This bursty transmission pattern can temporarily saturate network buffers, introducing queuing delays that do not accurately represent the true network congestion state. Since only the first packet of a burst is unaffected by the queuing delay induced by its subsequent packets, its RTT serves as a more reliable indicator of the underlying network conditions. 

\vspace{0.5em}
\noindent \textbf{Frame-Level Bandwidth-Delay Product.}  
With the estimated frame-level bandwidth and delay, we can further derive the frame-level bandwidth-delay product (BDP), leveraging similar principles from the classic congestion control model~\cite{cardwell2017bbr}:  
\begin{equation}
    \hat{BDP} = \text{avg}(B) \times \min(D),
    \vspace{-1mm}
\end{equation}

\noindent where $\text{avg}(B)$ denotes the average frame-level bandwidth, reflecting the actual available network capacity, and $\min(D)$ represents the minimum observed frame-level delay, which approximates the round-trip propagation time (RTprop).

\subsection{Congestion Detector}
\label{sec:design:cd}

To address challenge 2 and enable robust congestion detection at the frame level, \name\ uses the short-term trend of frame-level delay variations relative to the volume of in-flight data as its signal. Specifically, we compute the gradient of frame-level delay with respect to the number of bytes in flight. If this gradient exceeds a predefined threshold, congestion is inferred, as the gradient captures how queuing delay increases when the sender injects more data into the network. In the remainder of this section, we first justify the choice of this congestion signal through an experiment. We then describe how it is mapped to the cwnd scaling factor, and finally define the bandwidth value reported to the application layer.

\noindent \textbf{Rationale for the Congestion Signal.}
To compare the accuracy of our proposed congestion signal against several commonly used detection methods, we generated a synthetic trace. 
We assumed a fixed network configuration with a bandwidth of 2~Mbps and a round-trip propagation delay (RTprop) of 25~ms, resulting in a bandwidth-delay product (BDP) of 50~Kb. 
A series of in-flight data values were then generated to simulate the bitrate fluctuations of an encoder, and the RTT of each packet was computed as follows:
\begin{equation}
\text{RTT} =
\begin{cases} 
\text{RTprop} + \frac{\text{inflight} - \text{BDP}}{\text{Bandwidth}}, & \text{if inflight} > \text{BDP}, \\
\text{RTprop}, & \text{otherwise.}
\end{cases}
\label{eq:rtt}
\end{equation}

\noindent This formulation follows a typical queuing delay model: RTT remains constant at the propagation delay when the in-flight data volume is within the network capacity, but increases linearly once the BDP is exceeded. Based on this model, we define the ground-truth congestion state as any moment when the RTT exceeds RTprop. 

We compared our proposed inflight-based delay gradient signal, \(S(D, \text{inflight}) = \partial D / \partial \text{inflight}\), with two commonly used alternatives: the MinRTT-based queuing signal, \(S(D_{\min}) = D - D_{\min}\), and the time-based delay gradient, \(S(D, t) = \partial D / \partial t\). All methods utilized a consistent 5-second observation window for fair comparison. 

As shown in Figure~\ref{fig:design_cd}, the highlighted regions indicate time intervals where each method correctly or incorrectly identifies congestion. For the other two signals, errors occur after the queue has built up, whereas our inflight-based delay gradient consistently achieves substantially higher overall accuracy than the baselines.

\noindent \textbf{Mapping Congestion Signal to Window Scaling.}  
To control the sending window based on the detected congestion, we set the initial value of \(\gamma\) to 1. When congestion is detected, \(\gamma\) is reduced by multiplying it by 0.95 to proactively decrease the congestion window size and mitigate potential queue buildup: 
\begin{equation}
\gamma_{t+1} = \gamma_t \times 0.95,
\end{equation}

\noindent If no congestion is detected, $\gamma$ is reset to 1, allowing the cwnd to return to its full size and fully utilize the available bandwidth. 

\noindent \textbf{Congestion-Adjusted Bandwidth Reporting.}  
To provide the application layer with a more accurate view of available network capacity, \name\ reports a congestion-adjusted bandwidth to the application layer. The reported target bitrate is computed as: 
\begin{equation}
\text{target bitrate} = \gamma \times \text{avg}(B),
\end{equation}
where \(\gamma\) is the window-scaling factor derived from congestion detection and \(\text{avg}(B)\) is the average frame-level bandwidth.

\subsection{Bursting Length Controller}  
\label{sec:design:blc}

To address Challenge 3, \name\ introduces the bursting length controller, which aims to maximize the burst length without causing packet loss. We infer the network buffer length from the observed packet loss rate because we make the following observation: for deep buffers, packet loss is primarily caused by physical factors, manifesting as similar packet loss rates across frames of varying sizes. In contrast, for shallow buffers, packet loss is mainly due to buffer overflow, resulting in higher loss rates for larger frames. Specifically, we cut frames into a fixed 2 KB intervals, and the packet loss rate is computed separately for each interval. The smallest loss rate of the smallest interval, $L_0$, serves as an estimate of the physical loss rate. The buffer length estimate $M$ is updated every 5 seconds: 

\vspace{-3mm}
\begin{equation}
    M_t = 
    \begin{cases} 
        M_{t-1}-2\text{ KB}, & \text{if } L_i > L_0 + 0.1 \\[2mm]
        M_{t-1}+2\text{ KB}, & \text{otherwise}
    \end{cases}
    \label{equ:buffer_length_update}
    \vspace{-1mm}
\end{equation}

\noindent where $L_i$ refers to the loss rate of the current interval. If the loss rate of the current interval surpasses 10\% above $L_0$, $M$ is decremented. Otherwise, $M$ is incremented. If $M_t$ has already reached its minimum value and the loss rate is still above $L_0$, the algorithm falls back to a GCC-based approach. Once the buffer length $M$ is estimated, it is used to constrain the maximum frame length in periodic frame sending algorithm, mitigating the risk of buffer overflow.

\vspace{-2mm}
\section{Evaluations}
\label{sec:evaluation}


In this section, we evaluated \name~through five key aspects: 
Large-scale A/B testing in real-world scenarios (\S\ref{sec:eval_abtest});
Robustness under video bitrate undershooting conditions (\S\ref{sec:eval_undershoot});
Performance in weak network environments (\S\ref{sec:eval_weaknet});
Component analysis to isolate and understand individual contributions (\S\ref{sec:eval_comp}); and
Convergence behavior when competing with background traffic (\S\ref{sec:eval_converge}).

\noindent \textbf{Setup:}
Our large-scale experiments were conducted on one of the world’s largest video streaming platforms, which serves over 250 million users and 2 billion sessions across more than 150 countries. Due to the caution required for deploying new algorithms in a production environment, our large-scale A/B testing was limited to comparisons against the platform’s existing congestion control algorithm, which is a BBR-based variant. All other experiments were performed in a self-built software simulation environment. 

\noindent \textbf{Baselines:}
In the simulation experiments, \name~is compared against six baseline algorithms: GCC~\cite{carlucci2017congestion}, BBR~\cite{cardwell2017bbr}, Copa~\cite{arun2018copa}, Pudica~\cite{wang2024pudica}, SQP~\cite{ray2022sqp}, and Salsify~\cite{fouladi2018salsify}.
Due to Salsify’s requirement for a functional encoder, it could not be integrated into our streaming system. Thus, we evaluated it using its open-source implementation. Because of the substantial differences in processing pipelines, Salsify’s frame delay is not directly comparable and therefore omitted from our results. Moreover, as Salsify lacks an explicit bitrate target, we excluded it from the undershooting experiments.

\noindent \textbf{Metrics:} 
We evaluate \name~using four key metrics: \emph{Media Bitrate}, \emph{Frame Delay}, \emph{Stalling Ratio}, and \emph{Bandwidth Estimation Accuracy}.  
Media Bitrate quantifies the effective media transmission rate, reflecting both quality and efficiency.  
Frame Delay measures the end-to-end latency for a video frame.  
Stalling Ratio denotes the fraction of playback time spent in stalling, where stalling duration is defined as frame intervals exceeding 200~ms recorded by the server.  
Bandwidth Estimation Accuracy assesses how closely the bandwidth reported to the application layer matches the actual throughput, ensuring fair comparison across algorithms.

\subsection{Large-Scale Real-World A/B Test}
\label{sec:eval_abtest}

Overall, the large-scale A/B test demonstrates that \name~improves video quality and smoothness without compromising latency.

\noindent \textbf{Media bitrate.}  
As shown in Fig. \ref{fig:eva_ab_test_all}(ab), \name~consistently achieves higher media bitrate than the baseline.  
For Android users, it improves the average bitrate by 11.9\% in live streaming and 7.6\% in co-broadcasting.  
For iOS users, the gains reach 3.5\% and 14.4\%.

\noindent \textbf{Resolution.}  
Fig. \ref{fig:eva_ab_test_all}(ef) shows that \name~enhances video resolution by leveraging its higher bitrate.  
On Android, it increases the proportion of videos at or above 720P by 10.5\%.  
On iOS, the share of 1080P videos in live streaming rises from 24\% to 41\%, and the proportion of at least 576P videos in co-broadcasting grows by 27\%.

\noindent \textbf{Stalling ratio.}  
Fig. \ref{fig:eva_ab_test_all}(gh) shows that \name~achieves a lower stalling ratio overall.  
For Android, the difference is minor, while on iOS, the average stalling ratio in live streaming decreases by 14.1\%.  
In co-broadcasting, the average remains similar, but tail stalling events are notably reduced, improving user experience in extreme cases.

\noindent \textbf{Frame delay.}  
As shown in Fig. \ref{fig:eva_ab_test_all}(cd), \name~maintains frame delay comparable to the baseline across both Android and iOS, while achieving significant improvements in bitrate and stability.

\noindent \textbf{Padding bitrate.}  
The baseline maintaining an average padding bitrate of at least 5~Kbps in live streaming and 25~Kbps in co-broadcasting, while \name~ reduces padding bitrate to near zero because Padding occurs only for probing inactive connections.

\begin{figure}[t]
  \centering
  \begin{subfigure}[b]{0.48\linewidth}
    \includegraphics[width=\linewidth]{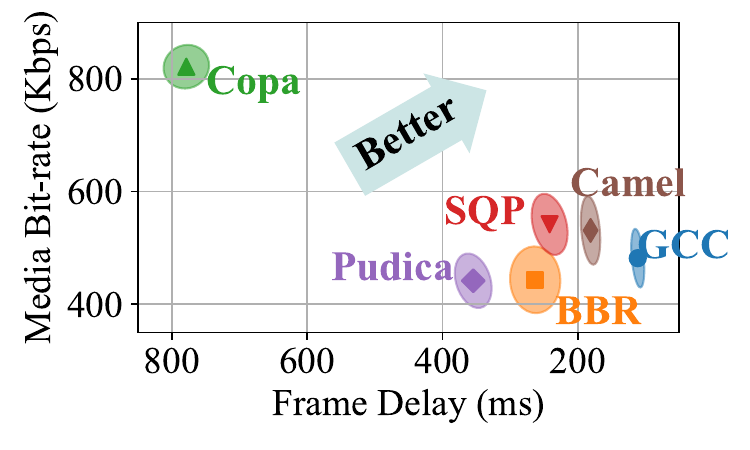}
    \vspace{-7mm}
    \caption{iOS}
    \label{fig:eva_applimited_overall_1}
  \end{subfigure}
  \hfill
  \begin{subfigure}[b]{0.48\linewidth}
    \includegraphics[width=\linewidth]{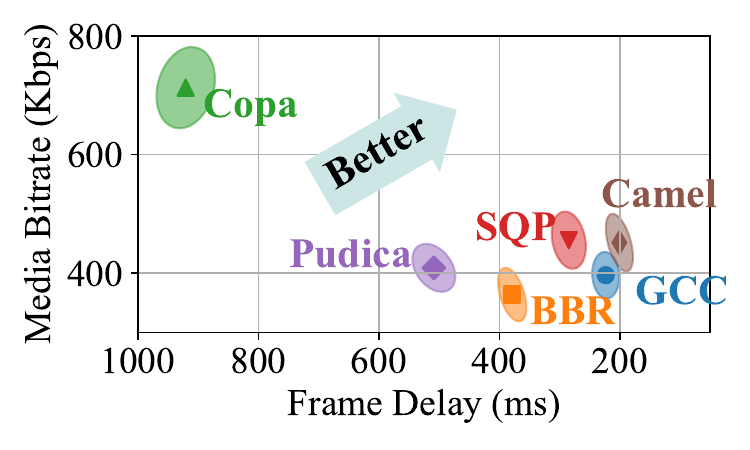}
    \vspace{-7mm}
    \caption{Android}
    \label{fig:eval_applimited_overall_2}
  \end{subfigure}
  \vspace{-4mm}
  \caption{Camel achieved a better trade-off between bitrate and frame delay in the simulation of undershooting scenario.}
  \label{fig:eval_overall}
  \vspace{-5mm}
\end{figure}

\begin{figure}[t]
  \centering
  \begin{subfigure}[b]{0.48\linewidth}
    \includegraphics[width=\linewidth]{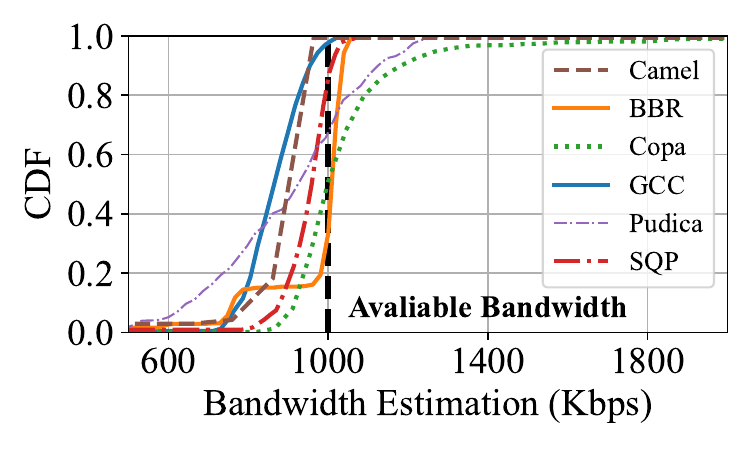}
    \vspace{-7mm}
    \caption{Without undershooting}
    \label{fig:eval_without_app}
  \end{subfigure}
  \hfill
  \begin{subfigure}[b]{0.48\linewidth}
    \includegraphics[width=\linewidth]{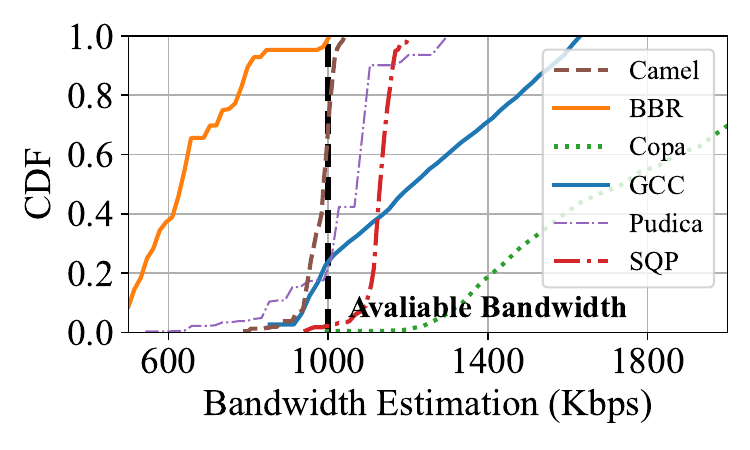}
    \vspace{-7mm}
    \caption{With undershooting}
    \label{fig:eval_with_app}
  \end{subfigure}
  \vspace{-4mm}
  \caption{Camel can accurately estimate bandwidth in video bitrate undershooting scenarios.}
  \label{fig:eval_app_change}
  \vspace{-5mm}
\end{figure}

\subsection{Robustness under undershooting}
\label{sec:eval_undershoot}

\noindent \textbf{Overall Performance.}
We evaluated the overall performance of all algorithms under video bitrate undershooting conditions.
Two sets of bitrate undershooting traces were simulated for iOS and Android, following the distribution of the co-broadcasting scenario shown in Figure~\ref{fig:back_app_limited_bw}(a).
The available bandwidth was capped at 1000~Kbps, and each algorithm ran for 150~seconds. 
Since no noticeable stalling was observed, we focused on media bitrate and frame delay.

As shown in Figure~\ref{fig:eval_overall}, \name~achieves higher bitrate than most baselines, except for Copa. 
In particular, it improves the average bitrate by 4.7\% on Android and 19.8\% on iOS compared to GCC.
This improvement stems from \name’s ability to accurately estimate available bandwidth even when the encoder output undershoots.
In contrast, BBR suffers from persistent bandwidth underestimation, while Pudica exhibits high volatility due to its assumption of stable frame sizes. 
Copa tends to overestimate bandwidth, resulting in excessive frame delays that are difficult to recover from promptly.

\noindent \textbf{Bandwidth Estimation.}
We further examined the accuracy of bandwidth estimation under undershooting conditions. 
The available bandwidth was fixed at 1000~Kbps, and each algorithm ran for 90~seconds. 
Between 30~s and 60~s, the encoding ratio was manually reduced to 60\% to emulate undershooting. 
We compared the estimated bandwidth during this period with that from the normal phase, as shown in Figure~\ref{fig:eval_app_change}. 
Among the baselines, only the frame-level algorithms SQP and Pudica are capable of estimating bandwidth under undershooting. BBR exhibits severe underestimation due to reduced receiving rates, while delay-based schemes such as GCC and Copa overestimate bandwidth when no delay buildup is observed.
In contrast, \name~achieves both higher accuracy and stability owing to its frame-level design, which remains robust even with varying frame sizes. 
Specifically, \name~achieves an average estimation accuracy of 98.9\%, improving upon GCC by 23.91\%.

\begin{figure}[t]
  \centering
  \includegraphics[width=1\linewidth]{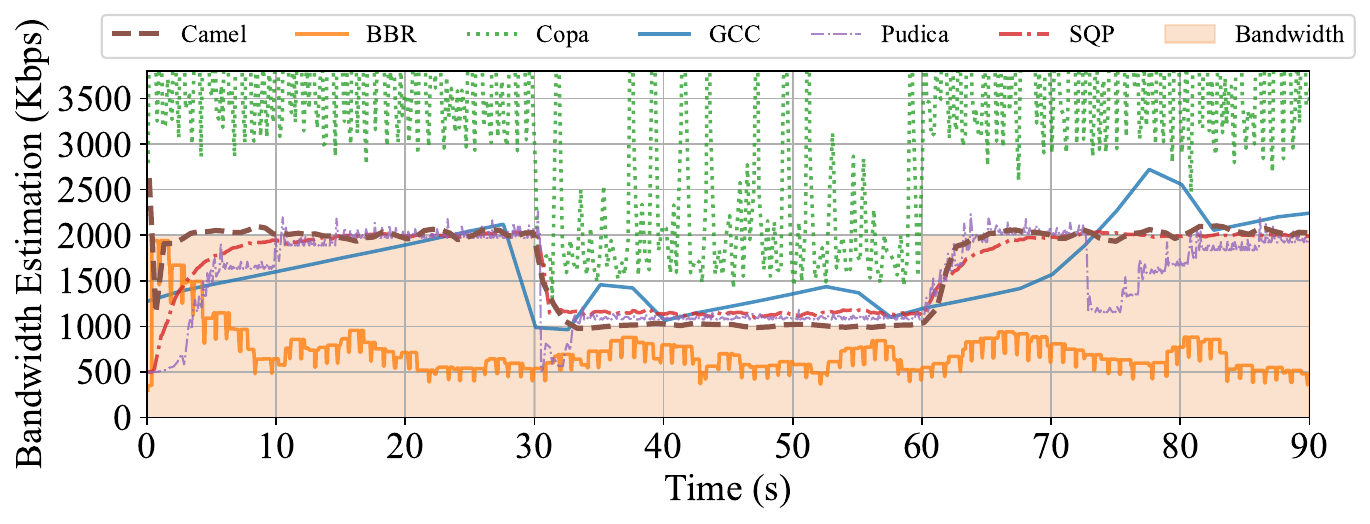}
  \vspace{-8mm}
  \caption{Camel can correctly follow bandwidth fluctuations under the video bitrate undershooting scenario. }
  \label{fig:eval_applimited_bw} 
  \vspace{-5mm}
\end{figure}

\noindent \textbf{Responsiveness.}
We evaluated the algorithms' responsiveness to bandwidth changes under undershooting. 
Each algorithm ran for 90~seconds: bandwidth was 2000~Kbps for the first 30~s, reduced to 1000~Kbps for the middle 30~s, and restored to 2000~Kbps for the final 30~s. 
Bandwidth estimates over time are shown in Figure~\ref{fig:eval_applimited_bw}. 
Frame-level algorithms, including \name, SQP, and Pudica, correctly tracked bandwidth fluctuations. 
BBR, however, remained low due to consistently reduced receiving rates under undershooting. 
Delay-based algorithms, GCC and Copa, overestimated bandwidth and thus responded poorly to actual network changes.

\subsection{Performance in weak network}
\label{sec:eval_weaknet}

\noindent \textbf{Performance under real-world 4G/5G traces.}
We evaluated \name~on 4 real-world LTE traces from the public dataset~\cite{lumos5g_imc20}, covering diverse network types and mobility conditions. 
Table~\ref{table:eva_lte_trace} shows that \name~reduces stalling ratio by 13\%--49\% compared to other frame-level methods (SQP, Pudica, Salsify). 
Salsify suffers from low frame rate due to high encoder resource usage, Pudica assumes constant frame sizes, and SQP converges slowly, causing stalls during sudden bandwidth drops. 
Compared to GCC, \name~achieves up to 52\% higher bitrate on 4G traces, as its packet-train-based design quickly reacts to bandwidth increases. 
Although BBR and Copa achieve high throughput and low stalling, \name~reduces frame delay by 12\%--17\%, avoiding periodic buffer-filling overheads.

\noindent \textbf{Performance against network jitter.}
Figure~\ref{fig:eva_jitter_perf} shows \name~under network jitter with 1000 kbps bandwidth. 
BBR, SQP, and Pudica rely on minRTT probing, misinterpreting transient jitter as persistent congestion, causing bandwidth underestimation, insufficient FEC transmission, severe stalling, and high frame delay. 
Copa mitigates jitter by filtering extreme RTTs, achieving higher bitrate than minRTT methods, but its conservative updates delay congestion resolution, leading to persistent stalls. 
GCC uses delay gradients and maintains low frame delay with minimal stalling. 
\name~further improves robustness by computing delay gradients over inflight-ordered frames instead of time-ordered packets, reducing jitter impact and achieving bitrate up to 94.9\% higher than GCC.


\noindent \textbf{Performance against packet loss.}
Figure~\ref{fig:eva_post_bottle_loss_perf} shows performance under packet loss with a 1000 Kbps bottleneck. 
As loss rate increases, all algorithms see bitrate drops due to reduced effective capacity. 
GCC and BBR severely underestimate bandwidth at high loss rates, impairing retransmission and FEC efficiency, and causing more rebuffering. 
SQP and Pudica overestimate bandwidth under high loss, also increasing rebuffering. 
At 50\% loss, their stalling ratios are 4.53\% and 2.62\% higher than \name, respectively.

\noindent \textbf{Performance under shallow buffer cases.} 
Figure~\ref{fig:eva_shallow_buffer_qoe} compares \name~with baseline algorithms for buffer sizes from 2KB to 10KB. 
At 2KB, SQP and Pudica lack packet samples for accurate bandwidth estimation, while Copa fails to detect congestion due to low queuing delay. 
These algorithms suffer persistent overshooting, causing high stalling and longer delay; packet loss further reduces media bitrate. 
BBR and GCC perform well by trading off bitrate and delay. 
\name~detects shallow buffers and falls back to GCC, achieving comparable performance and outperforming other frame-level methods. 
As buffer length grows, GCC degrades due to frequent mode switching, while SQP and Pudica only start estimating bandwidth accurately above 4KB and 6KB, respectively. 
\name~still provides superior QoE by optimally adjusting frame length.

\subsection{Component Analysis}
\label{sec:eval_comp}

\begin{figure*}[t]
  \centering
  \begin{minipage}[b]{0.24\linewidth}
    \includegraphics[width=\linewidth]{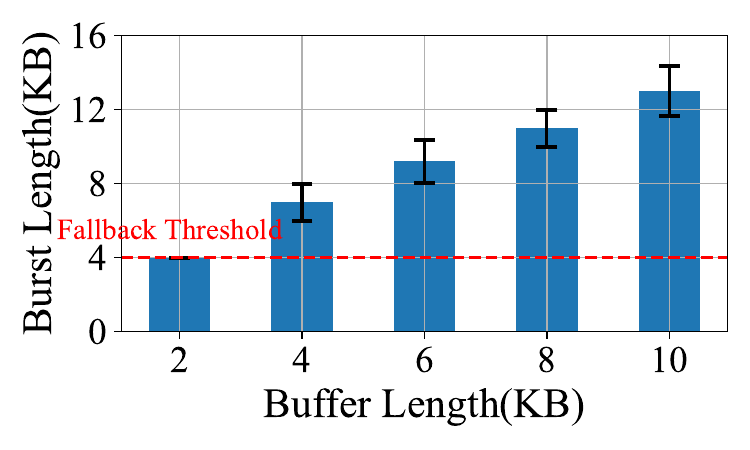}
    \vspace{-8mm}
    \caption{\name's determined burst length for different network buffer lengths.}
    \label{fig:design_buffer_valid} 
  \end{minipage}
  \begin{minipage}[b]{0.75\linewidth}
      \begin{subfigure}[b]{0.32\linewidth}
        \includegraphics[width=\linewidth]{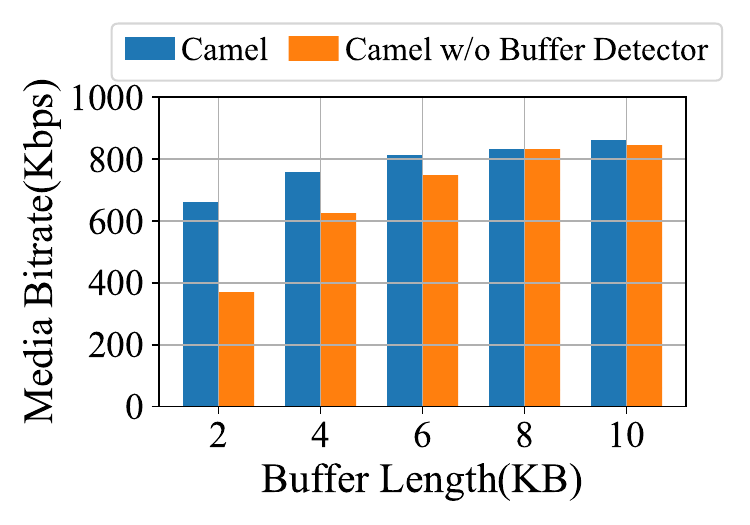}
        \vspace{-6mm}
        \caption{Media bitrate}
        \label{fig:eva_buffer_bitrate}
      \end{subfigure}
      \hfill
      \begin{subfigure}[b]{0.32\linewidth}
        \includegraphics[width=\linewidth]{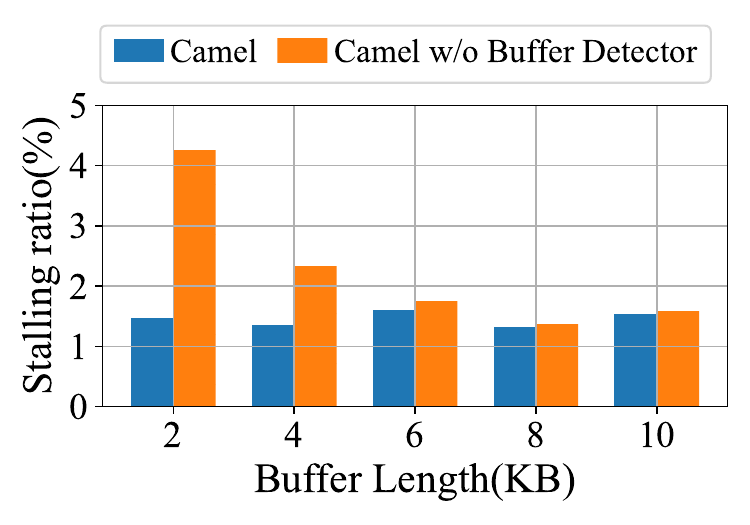}
        \vspace{-6mm}
        \caption{Stalling ratio}
        \label{fig:eva_buffer_stall}
      \end{subfigure}
      \hfill
      \begin{subfigure}[b]{0.32\linewidth}
        \includegraphics[width=\linewidth]{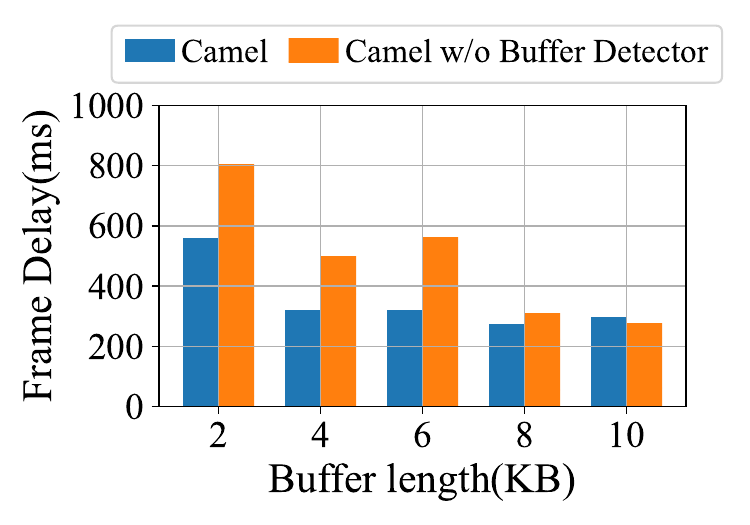}
        \vspace{-6mm}
        \caption{Frame delay}
        \label{fig:eva_buffer_delay}
      \end{subfigure}
      \vspace{-4mm}
      \caption{\name's QoE improvement with Bursting Length Controller~against shallow buffer.}
      \label{fig:eva_buffer_effect}
  \end{minipage}
  \vspace{-2mm}
\end{figure*}

\noindent \textbf{Bursting Length Controller.} 
The effectiveness of the Bursting Length Controller is shown in Figure~\ref{fig:design_buffer_valid}. 
For very shallow buffers (2KB), where packet train-based methods fail, the controller reduces burst length, enabling \name~to fall back to GCC. 
As buffer length increases, burst length scales accordingly, maximizing samples for \sampler~while avoiding buffer overflow. 
At 10KB, bursts exceed 12KB, allowing most large frames to be sent fully, reducing frame delay and improving bandwidth estimation.

Figure~\ref{fig:eva_buffer_effect} compares Bursting Length Controller with a fixed 20KB baseline. 
In a 2KB buffer, it achieves 1.79$\times$ media bitrate and reduces stalling by 67\%. 
For 4–6KB buffers, it prevents unnecessary packet loss, reduces retransmissions and FEC, significantly improves QoE.

\subsection{Convergence with Background Traffic}
\label{sec:eval_converge}

We evaluated \name's congestion detection in multi-stream scenarios, using a common bandwidth of 1000 Kbps. 
Results showes that it achieves good fairness with various types of background traffic.

\begin{figure}[t]
  \centering
  \begin{subfigure}[b]{0.49\linewidth}
    \includegraphics[width=\linewidth]{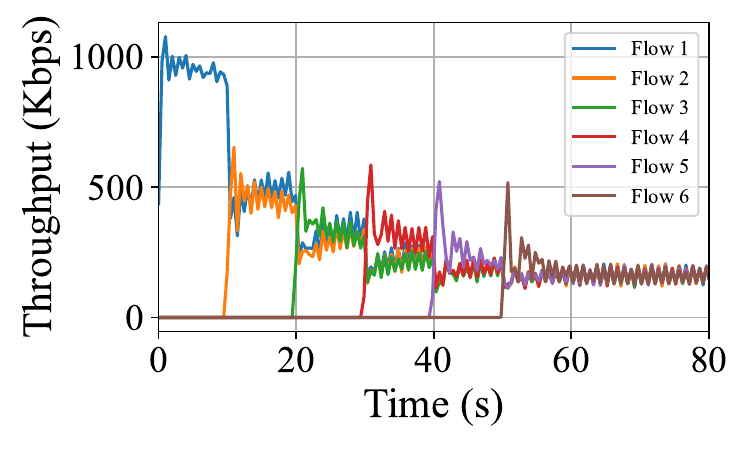}
    \label{fig:eval_converge_multiflow_bw}
  \end{subfigure}
  \hfill
  \begin{subfigure}[b]{0.49\linewidth}
    \includegraphics[width=\linewidth]{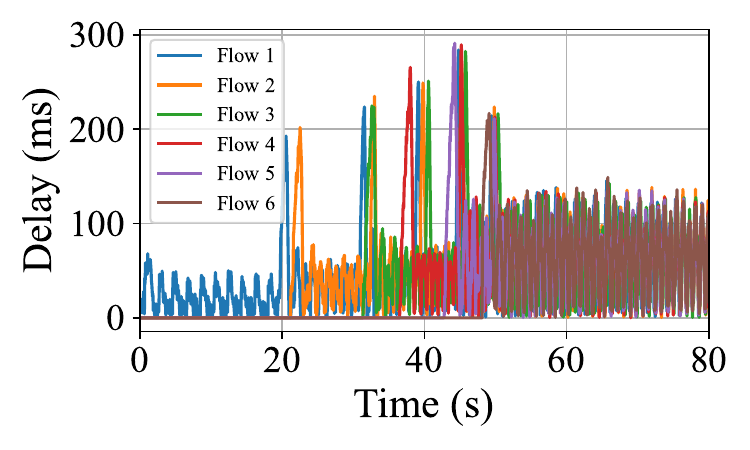}
    \label{fig:eval_converge_mutiflow_delay}
  \end{subfigure}
  \vspace{-9mm}
  \caption{Camel can achieve good inter-flow fairness.}
  \label{fig:eval_converge_mutiflow}
  \vspace{-4mm}
\end{figure}

\noindent \textbf{Inter-flow fairness.} 
We evaluated the fairness among multiple \name~flows with the same RTT.
A new \name~flow was introduced every 10 seconds.
As shown in Figure~\ref{fig:eval_converge_mutiflow}, \name~successfully achieved fair bandwidth allocation and convergent latency.

\begin{figure}[t]
  \centering
  \begin{subfigure}[b]{0.49\linewidth}
    \includegraphics[width=\linewidth]{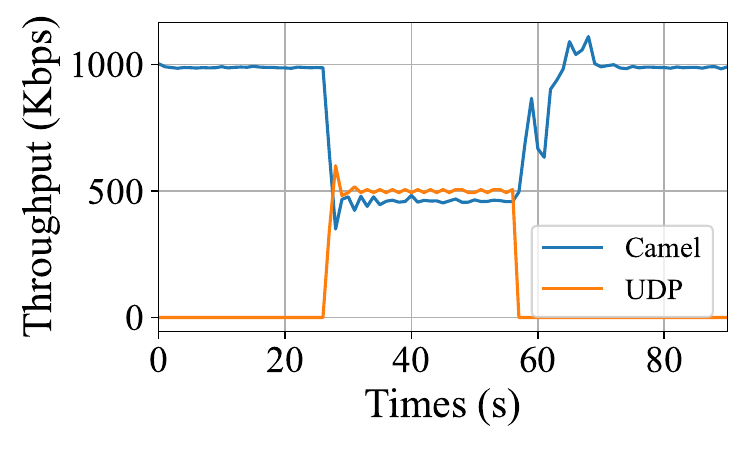}
    \label{fig:eval_converge_udp_bw}
  \end{subfigure}
  \hfill
  \begin{subfigure}[b]{0.49\linewidth}
    \includegraphics[width=\linewidth]{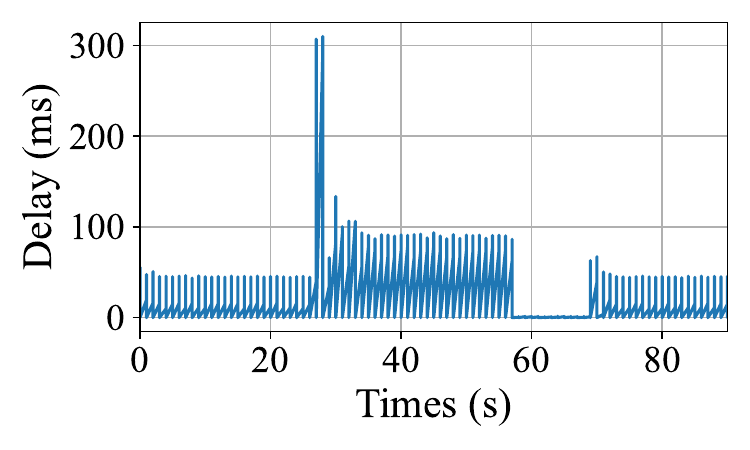}
    \label{fig:eval_converge_udp_delay}
  \end{subfigure}
  \vspace{-9mm}
  \caption{Coexisting with UDP.}
  \label{fig:eval_converge_udp}
  \vspace{-6mm}
\end{figure}

\noindent \textbf{Competitiveness with Inelastic Traffic.} 
We evaluated \name's ability to coexist with inelastic traffic, such as UDP, ensuring that it does not continuously compete for bandwidth unnecessarily. The experiment ran for 90 seconds, during which an UDP flow joint in. We collected the throughput and packet one-way delay. As shown in Figure~\ref{fig:eval_converge_udp}, \name~effectively shared bandwidth with UDP while maintaining stable and controlled latency.

\noindent \textbf{Competitiveness with Elastic Traffic.} 
We then evaluated \name's ability to coexist with elastic traffic like GCC, ensuring that it neither starves nor being starved when competing with low-latency algorithms. \name~ran for 90 seconds, during which a GCC flow was induced. Figure~\ref{fig:eval_converge_gcc} shows the traces of throughput and RTT of the two flows, indicating that \name~successfully shared bandwidth with GCC while achieving controllable latency.
When coexisting with buffer fillers like TCP Cubic, \name~does not aggressively compete for bandwidth, as excessive latency negatively impacts the experience of LLL streaming. While \name\ is able to maintain a fair share of bandwidth under shallow network buffers, allowing TCP to quickly gain bandwidth, complete its data transmission, and exit the network can be an effective strategy in deep buffer scenarios.
Figure~\ref{fig:eval_converge_tcp} presents an experiment where a TCP CUBIC flow was introduced during \name's video stream transmission. The results show that when the network buffer is set to 500 KB, \name~maintains a minimal transmission rate. However, when the network buffer is decreased to 100 KB, \name~achieves roughly equal bandwidth sharing with TCP CUBIC.

\section{Related Works About Bandwidth Prediction in Undershooting Condition}
\label{sec:related}

\textbf{Application-limited.}
Sammy~\cite{spang2023sammy} is a new adaptive bitrate (ABR) algorithm designed to smooth the video traffic, making it more friendly to internet applications. 
However, when traffic is smoothed, throughput-based algorithm struggle to accurately estimate bandwidth, placing Sammy in the same "Application-Limited" condition as our approach. 
To mitigate this, Sammy over-transmits traffic during initialization and subsequently adjusts the level of over-transmission based on specific algorithmic requirements. 
However, if the underlying CCA can effectively handle application-limited conditions, Sammy would have greater flexibility in selecting ABR algorithms while maintaining smooth traffic.

\textbf{Adaptive Bitrate and Frame Rate.}
ABR and frame rate adaption are hot topics for LLS.
Fugu~\cite{fugu} leverages a deep neural network trained on real-world data for robust throughput prediction.
Sensei~\cite{sensei} improves ABR algorithms by evaluating users' quality sensitivity to different parts of the video. 
SODA~\cite{chen2024soda} optimizes three key metrics: bitrate switching, bitrate, and stalling, while reducing algorithmic complexity to enhance deployability.
ARTEMIS~\cite{ARTEMIS} dynamically configures the bitrate ladder for live streaming to adapt to time-varying video content.  
AFR~\cite{meng2023enabling} introduces adaptive frame rate control to improve the frame rate and resolution of video streaming.
If CCAs can provide stable and accurate bandwidth estimation even under application-limited conditions, ABR and frame rate adaptation can be further optimized for improved performance.

\textbf{Transport Layer Optimization.} 
Beyond the design of CCA, there are several other optimization approaches at the transport layer.
AUGUR~\cite{AUGUR} leverages multi-path transport over cellular networks to reduce long tail latency caused by Wi-Fi fluctuation in cloud gaming.
Converge~\cite{dhawaskar2023converge} enhances multi-path algorithms by incorporating video frame information, improving frame transmission integrity.
Zhuge~\cite{meng2022achieving} accelerates the response of CCAs by proactively blocking ACKs on the access point (AP) based on the status of network queue.
Hairpin~\cite{hairpin} introduces a loss recovery mechanism that optimizes the combination of retransmissions and forward error correction (FEC) to meet deadlines while conserving bandwidth for edge-based interactive streaming.

\textbf{Encoding and Transmission Interconnection.}
Several approaches have explored the coordination between encoding and transport layers to achieve better performance, often at the cost of increased deployment complexity.
Swift~\cite{swift} employs layered coding with neural video codecs to mitigate ABR issues caused by inaccurate network capacity estimations. BurstRTC~\cite{jia2024burstrtc,jiatackling} predicts frame delay and QoE by modeling the distribution of frame sizes. 
Salsify~\cite{fouladi2018salsify} enhances encoder efficiency by integrating it with the transport layer through a custom encoder.
\section{Conclusion}
\label{sec:conclusion}

This paper addresses a critical bottleneck in designing upstream congestion control algorithms for existing low-latency live streaming systems, where video bitrate undershooting is prevalent. 
We propose \name, a novel frame-level congestion control algorithm that mitigates the dependence on large or continuous data. 
Experimental results demonstrate that \name~provides accurate bandwidth estimation across diverse network conditions, while significantly improving media bitrate and reducing rebuffering in real-world deployments. 
This work does not pose any ethical concerns.
This work was sponsored by the NSFC grant(62431017), ByteDance Grant(CT20241126107484). We gratefully acknowledge the support of Key Laboratory of Intelligent Press Media Technology.
The corresponding author is Xinggong Zhang.

\bibliographystyle{ACM-Reference-Format}
\bibliography{sample-base}

\appendix

\begin{figure*}
    \begin{minipage}[b]{0.48\linewidth}
      \begin{subfigure}[b]{0.49\linewidth}
        \includegraphics[width=\linewidth]{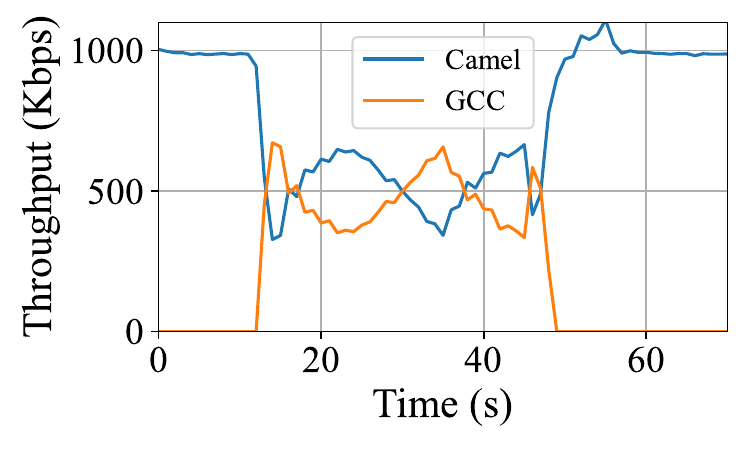}
        \label{fig:eval_converge_gcc_bw}
      \end{subfigure}
      \hfill
      \begin{subfigure}[b]{0.49\linewidth}
        \includegraphics[width=\linewidth]{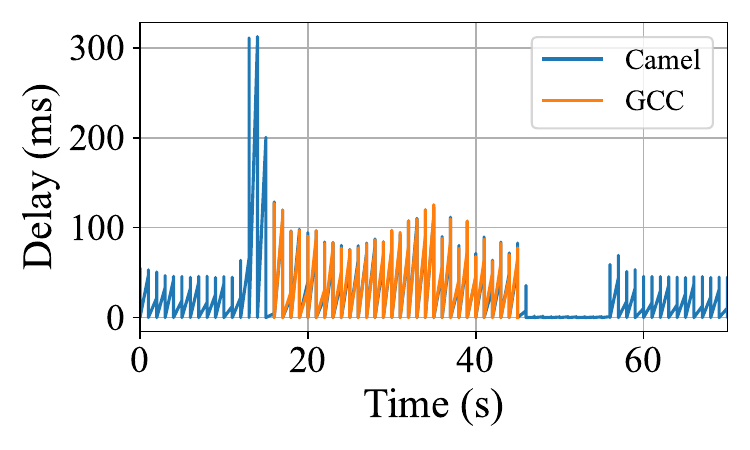}
        \label{fig:eval_converge_gcc_delay}
      \end{subfigure}
      \vspace{-7mm}
      \caption{Coexisting with GCC.}
      \label{fig:eval_converge_gcc}
      \vspace{5mm}
    \end{minipage}
    \hfill
    \begin{minipage}[b]{0.48\linewidth}
      \begin{subfigure}[b]{0.49\linewidth}
        \includegraphics[width=\linewidth]{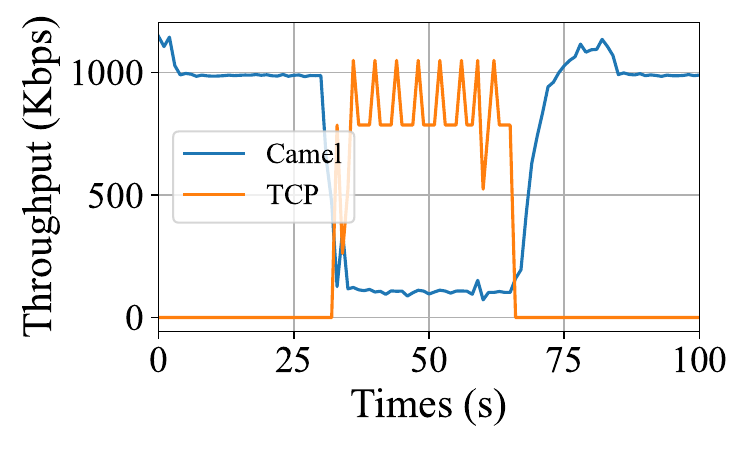}
        \vspace{-5mm}
        \caption{$buffer=500KB$.}
        \label{fig:eval_converge_tcp_bw_200ms}
      \end{subfigure}
      \hfill
      \begin{subfigure}[b]{0.49\linewidth}
        \includegraphics[width=\linewidth]{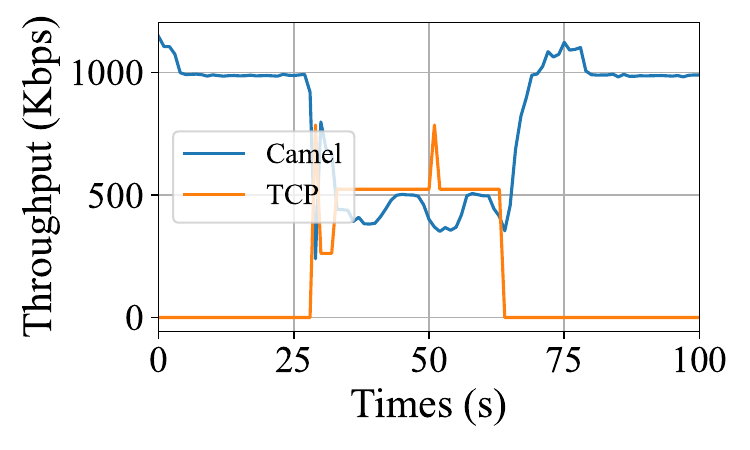}
        \vspace{-5mm}
        \caption{$buffer=100KB$.}
        \label{fig:eval_converge_tcp_bw_500ms}
      \end{subfigure}
      \vspace{-3mm}
      \caption{Coexisting with TCP.}
      \label{fig:eval_converge_tcp}
    \end{minipage}
\vspace{4mm}
\end{figure*}

\begin{table*}[t]
\centering
\caption{Comparison of QoE over real-world LTE traces.}
\vspace{-4mm}
\resizebox{2\columnwidth}{!}{%
\begin{tabular}{@{}c|ccc|ccc|ccc|ccc@{}}
\toprule
        & \multicolumn{3}{c|}{4G\_walking\_50167} & \multicolumn{3}{c|}{4G\_driving\_50187} & \multicolumn{3}{c|}{5G\_walking\_90} & \multicolumn{3}{c}{5G\_walking\_109} \\ \midrule
        & \begin{tabular}[c]{@{}c@{}}Bitrate\\ (Kbps)\end{tabular} 
        & \begin{tabular}[c]{@{}c@{}}Delay\\ (ms)\end{tabular} 
        & \begin{tabular}[c]{@{}c@{}}Stall\\ (\%)\end{tabular} 
        & \begin{tabular}[c]{@{}c@{}}Bitrate\\ (Kbps)\end{tabular} 
        & \begin{tabular}[c]{@{}c@{}}Delay\\ (ms)\end{tabular} 
        & \begin{tabular}[c]{@{}c@{}}Stall\\ (\%)\end{tabular} 
        & \begin{tabular}[c]{@{}c@{}}Bitrate\\ (Kbps)\end{tabular} 
        & \begin{tabular}[c]{@{}c@{}}Delay\\ (ms)\end{tabular} 
        & \begin{tabular}[c]{@{}c@{}}Stall\\ (\%)\end{tabular} 
        & \begin{tabular}[c]{@{}c@{}}Bitrate\\ (Kbps)\end{tabular} 
        & \begin{tabular}[c]{@{}c@{}}Delay\\ (ms)\end{tabular} 
        & \begin{tabular}[c]{@{}c@{}}Stall\\ (\%)\end{tabular} \\ \midrule
Camel   & 1606.61 & 1014.57 & 6.35 & 2010.09 & 183.69 & 0.30 & 2030.23 & 135.34 & 1.42 & 1981.05 & 546.51 & 4.99 \\
Copa    & 1697.66 & 997.89  & 6.78 & 2035.74 & 268.16 & 0.00 & 2042.70 & 185.39 & 1.36 & 2052.34 & 652.03 & 5.05 \\
BBR     & 1668.18 & 1167.29 & 10.92 & 2014.49 & 195.75 & 0.29 & 2028.53 & 213.41 & 1.34 & 2042.48 & 687.53 & 4.84 \\
GCC     & 1054.97 & 704.40  & 6.05 & 1624.72 & 139.36 & 1.02 & 2037.92 & 153.65 & 1.41 & 1958.70 & 606.71 & 5.40 \\
SQP     & 1419.80 & 914.53  & 9.72 & 1954.26 & 188.76 & 2.41 & 1979.38 & 187.53 & 2.67 & 1860.21 & 634.29 & 8.61 \\
Pudica  & 1535.89 & 1037.90 & 7.73 & 1930.93 & 275.97 & 1.20 & 1946.38 & 231.36 & 3.65 & 1884.94 & 603.07 & 9.03 \\
Salsify & 1504.33 & /       & 12.69 & 2041.93 & /       & 2.03 & 2059.46 & /       & 1.56 & 2025.78 & /       & 9.59 \\ 
\bottomrule
\end{tabular}%
}
\label{table:eva_lte_trace}
\end{table*}

\begin{figure*}[t]
      \centering
      \begin{subfigure}[b]{0.32\linewidth}
        \includegraphics[width=\linewidth]{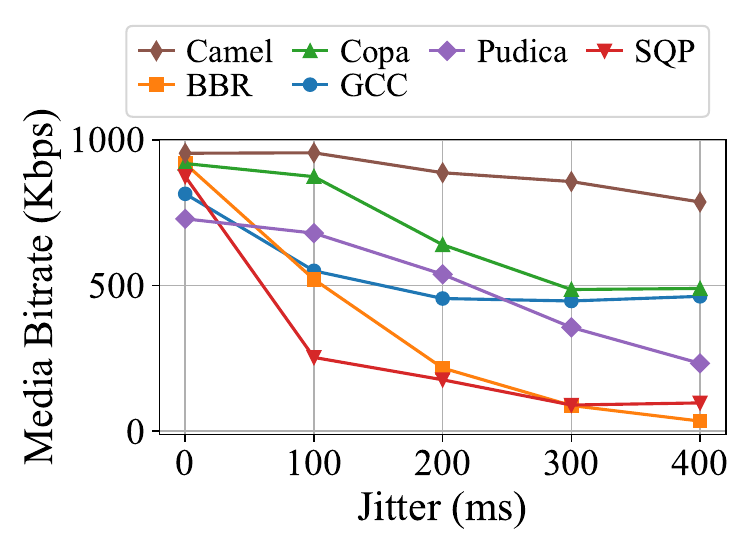}
        \vspace{-7mm}
        \caption{Media bitrate.}
        \label{fig:eva_jitter_rate}
      \end{subfigure}
      \hfill
      \begin{subfigure}[b]{0.32\linewidth}
        \includegraphics[width=\linewidth]{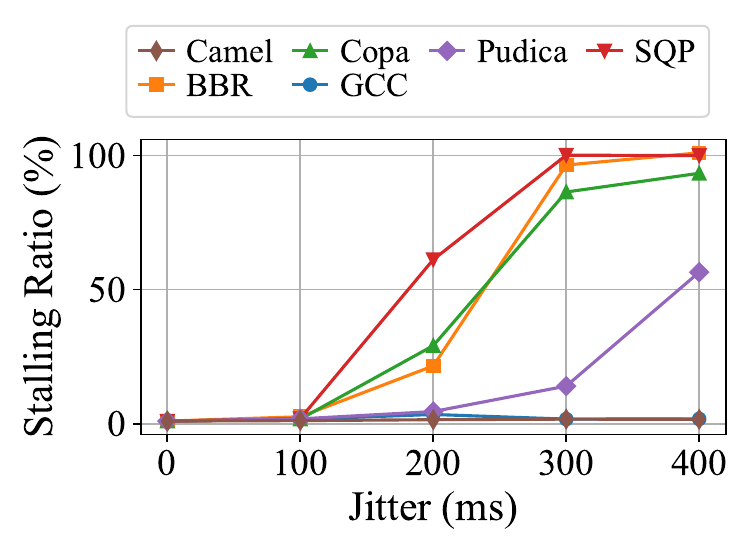}
        \vspace{-7mm}
        \caption{Stalling ratio.}
        \label{fig:eva_jitter_stall}
      \end{subfigure}
      \hfill
      \begin{subfigure}[b]{0.32\linewidth}
        \includegraphics[width=\linewidth]{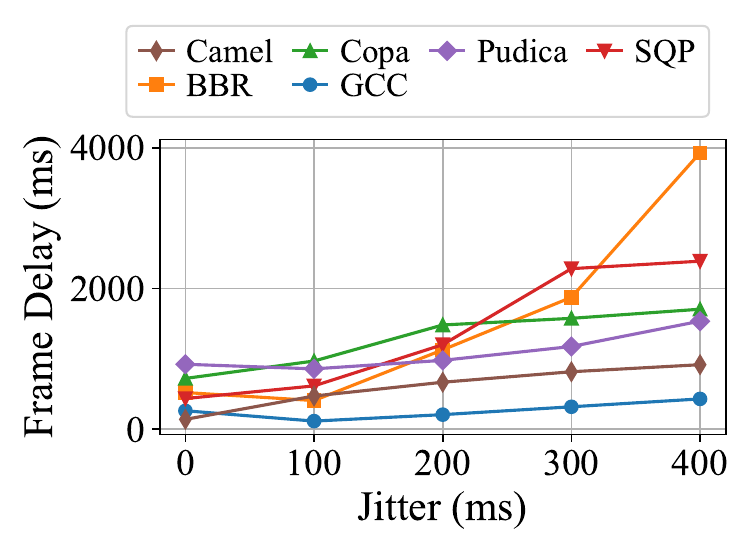}
        \vspace{-7mm}
        \caption{Frame Delay.}
        \label{fig:eva_jitter_delay}
      \end{subfigure}
    \vspace{-4mm}
    \caption{Performance under jitter.}
    \label{fig:eva_jitter_perf}
\end{figure*}

\begin{figure*}[t]
      \centering
      \begin{subfigure}[b]{0.32\linewidth}
        \includegraphics[width=\linewidth]{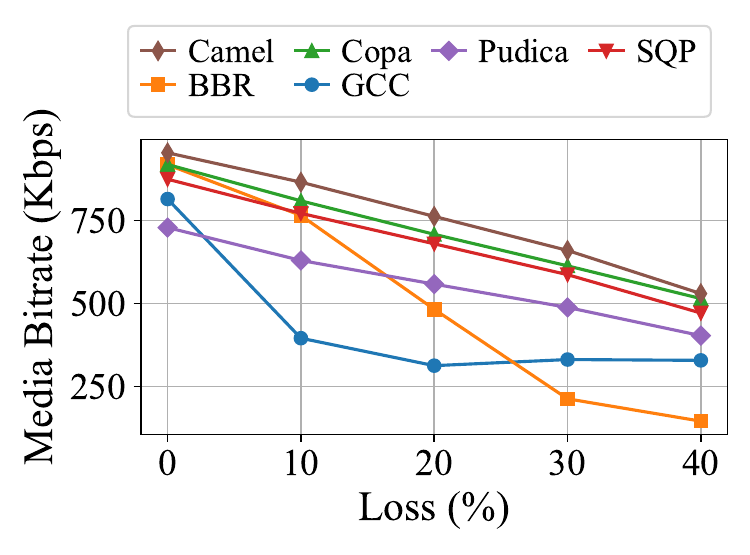}
        \vspace{-7mm}
        \caption{Media bitrate.}
        \label{fig:eva_post_bottle_loss_stall}
      \end{subfigure}
      \hfill
      \begin{subfigure}[b]{0.32\linewidth}
        \includegraphics[width=\linewidth]{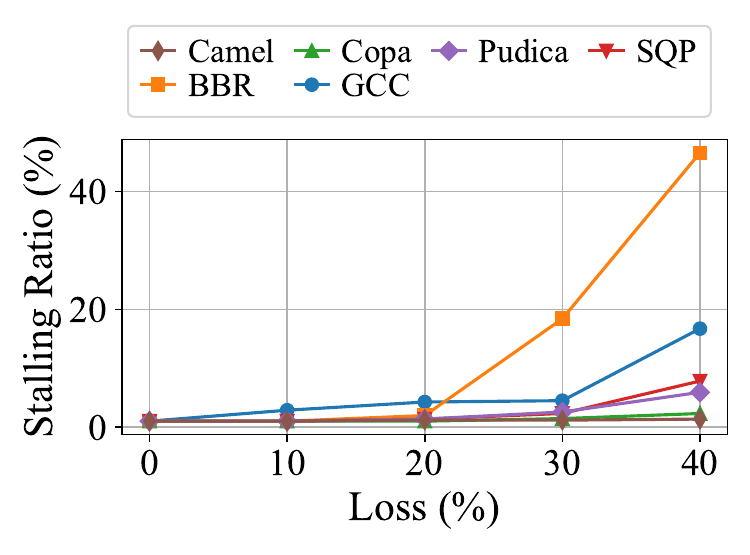}
        \vspace{-7mm}
        \caption{Stalling ratio.}
        \label{fig:eva_post_bottle_loss_bitrate}
      \end{subfigure}
      \hfill
      \begin{subfigure}[b]{0.32\linewidth}
        \includegraphics[width=\linewidth]{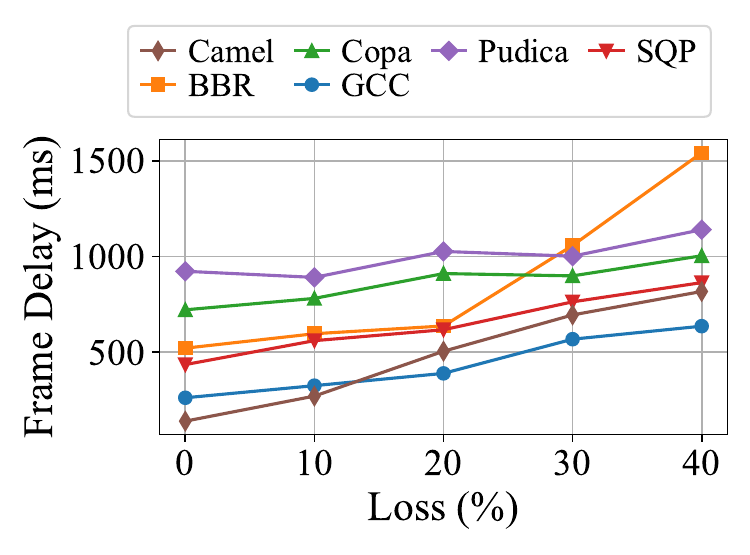}
        \vspace{-7mm}
        \caption{Frame Delay.}
        \label{fig:eva_post_bottle_loss_frame_delay}
      \end{subfigure}
    \vspace{-4mm}
    \caption{Performance under loss.}
    \label{fig:eva_post_bottle_loss_perf}
\end{figure*}

\begin{figure*}[t]
  \centering
  \begin{minipage}[b]{\linewidth}
      \begin{subfigure}[b]{0.32\linewidth}
        \includegraphics[width=\linewidth]{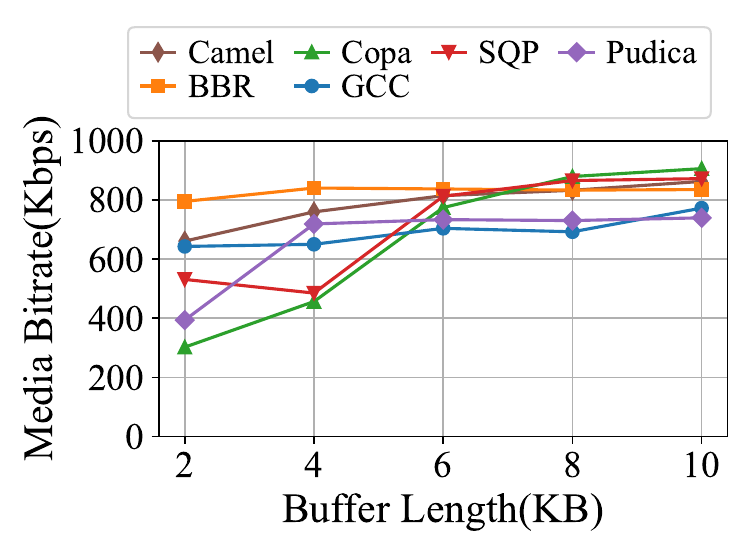}
        \vspace{-7mm}
        \caption{Media bitrate}
        \label{fig:eva_shallow_buffer_bitrate}
      \end{subfigure}
      \hfill
      \begin{subfigure}[b]{0.32\linewidth}
        \includegraphics[width=\linewidth]{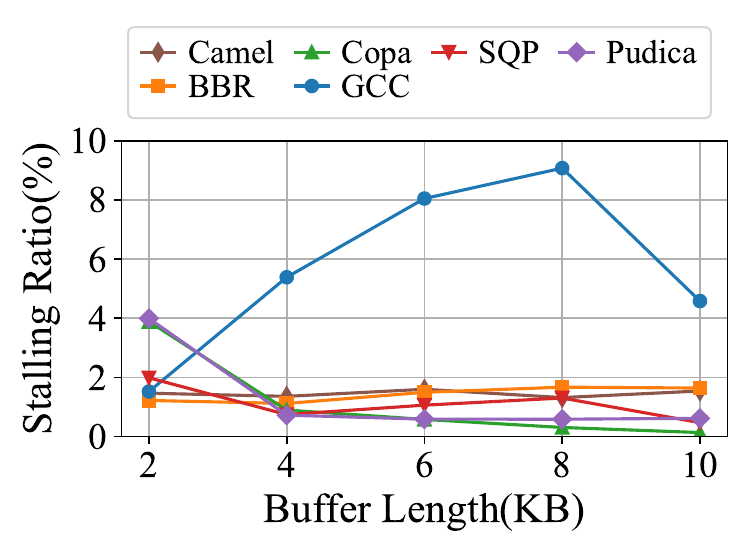}
        \vspace{-7mm}
        \caption{Stalling ratio}
        \label{fig:eva_shallow_buffer_stall}
      \end{subfigure}
      \hfill
      \begin{subfigure}[b]{0.32\linewidth}
        \includegraphics[width=\linewidth]{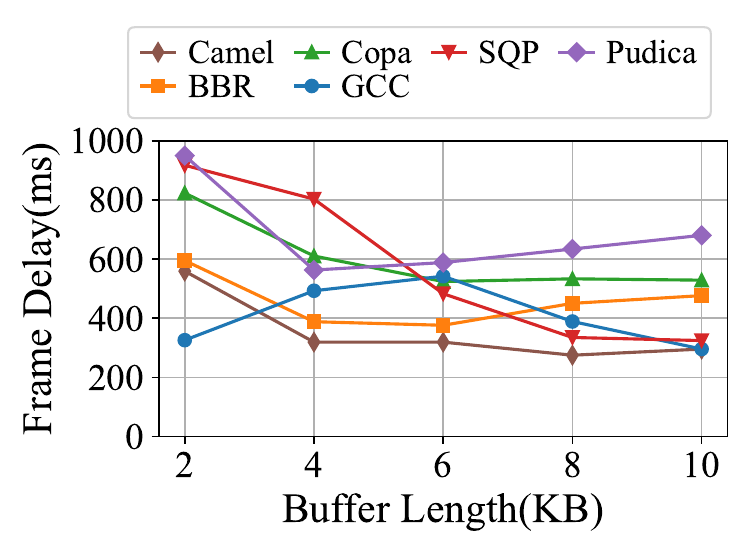}
        \vspace{-7mm}
        \caption{Frame delay}
        \label{fig:eva_shallow_buffer_delay}
      \end{subfigure}
      
  \end{minipage}
  \vspace{-8mm}
  \caption{Camel can maintain high bitrates, low stalling, and low frame delay under shallow buffer conditions.}
  \label{fig:eva_shallow_buffer_qoe}
\end{figure*}









\end{document}